\documentclass[structabstract]{aa}

\pdfoutput=1

\usepackage[fleqn]{amsmath}

\usepackage{graphicx}

\usepackage{txfonts}

\usepackage{natbib}
\bibpunct{(}{)}{;}{a}{}{,}

\usepackage[pdftex=true,colorlinks=true]{hyperref}
\usepackage[all]{hypcap}
\hypersetup{
pdfpagemode = {UseOutlines},
baseurl = {http://www.mpia.de/homes/kuiper},
pdftitle = {Fast and accurate frequency dependent radiation transport for hydrodynamics simulations in massive star
formation}, 
pdfauthor = {R.~Kuiper},
pdfkeywords = {radiative transfer - hydrodynamics - stars: formation - stars: circumstellar matter - accretion disks -
methods: numerical} 
}

\title{
Fast and accurate frequency dependent radiation transport\\ 
for hydrodynamics simulations in massive star formation
}
\titlerunning{Frequency dependent radiation transport for hydrodynamics simulations in massive star formation}

\author{ 
R.~Kuiper\inst{1,3} 
\and H.~Klahr\inst{1} 
\and C.~Dullemond\inst{1} 
\and W.~Kley\inst{2} 
\and T.~Henning\inst{1} 
}
\authorrunning{R.~Kuiper et al.}

\institute{ 
Max-Planck-Institut f\"ur Astronomie, 
Abt.~Planeten- und Sternentstehung,
K\"onigstuhl 17, 
D-69117 Heidelberg, 
Germany \\
\email{kuiper@mpia.de}
\and 
Universit\"at T\"ubingen,
Institut f\"ur Astronomie und Astrophysik, 
Abt.~Computational Physics, 
Auf der Mor\-gen\-stelle 10, 
D-72076 T\"ubingen, 
Germany 
\and 
Fellow of the 
International Max-Planck Research School for Astronomy and Cosmic Physics at the University of Heidelberg (IMPRS-HD) 
}

\date{ Received {\it date} / Accepted {\it date} }

\abstract{
Radiative feedback plays a crucial role in the formation of massive stars. 
The implementation of a fast and accurate description of the proceeding thermodynamics in pre-stellar cores
and evolving accretion disks is therefore a main effort in current hydrodynamics simulations. 
} 
{
We introduce our newly implemented three-dimensional frequency dependent radiation transport
algorithm for hydrodynamics simulations of spatial configurations with a dominant central source.
} 
{
The module combines the advantage of the speed of an approximate Flux Limited Diffusion (FLD) solver
in the one-temperature approach, which is valid in the static diffusion limit,
with the high accuracy of a frequency dependent first order ray-tracing routine. 
The ray-tracing routine especially compensates the introduced 
inaccuracies by standard approximate FLD solvers in transition regions from optically thin to thick and yields the
correct optical depths for the frequency dependent stellar irradiation. 
Both components of our module make use of realistic tabulated dust opacities. 
The module is parallelized for distributed memory machines based on the message passing interface 
standard.
We implemented the module in the 
three-dimensional 
high-order 
magneto-hydrodynamics code Pluto. 
} 
{
We prove the viability of the scheme in a standard radiation benchmark test compared to a full frequency dependent
Monte-Carlo based radiative transfer code. 
The setup includes a central star, 
a circumstellar flared disk,
as well as an envelope. 
The test is performed for different optical depths. 
Considering the frequency dependence of the stellar irradiation, 
the temperature distributions can be described
precisely in the optically thin, thick, and irradiated transition regions. 
Resulting radiative forces onto dust grains are reproduced with high accuracy. 
The achievable parallel speedup of the method imposes no restriction on further radiative (magneto-) hydrodynamics
simulations. 
} 
{
The proposed approximate radiation transport method enables frequency dependent radiation hydrodynamics studies of the
evolution of pre-stellar cores and circumstellar accretion disks around an evolving massive star in a highly efficient
and accurate manner. 
}


\begin{document}

\maketitle

\section{Introduction}
\label{sect:introduction}
Massive stars are of importance for a wide range of astrophysical problems. 
In spite of the fact that they are both rare and short-living, 
they represent the major source of radiation energy in their stellar clusters.
Therefore, 
they act as valuable tracers of star formation rates in distant galaxies. 
Additionally, 
the radiation emitted during a massive star's lifetime influences the
surroundings through various interactions 
such as heating and ionization of gas, 
evaporation of dust, 
and radiative forces leading to powerful stellar winds and outflows.
Finally, when a massive star dies, it enriches its neighborhood with heavy elements.
In this sense, 
massive stars are the main drivers of the 
morphological, 
dynamical, 
and chemical evolution of their complex environments 
\protect\citep{Zinnecker:2007p752, McKee:2007p848, Henning:2008p1955}.

However, know\-ledge about their formation is rather poor compared to the case of low-mass star formation.
Observationally, this is mostly due to their
larger average distances and the fact that they are deeply embedded in dense, opaque cores
especially during their early evolutionary phases.
Also, 
their short lifetime, 
rareness, 
and complex environment pose difficulties for detailed observations.
Nevertheless, 
current observational results, 
e.g.~from the Spitzer Space Telescope Survey and millimeter interferometry, 
support the assumption that the basic concepts of star formation, 
including 
the collapse of an unstable gas and dust core, 
the formation of both bipolar outflows and jets, 
as well as accretion disks, 
are possibly valid throughout the whole range of stellar masses 
\citep[e.g.][]{Bally:2008p1950, Bik:2008p1946, Kumar:2008p1936, Steinacker:2008p2593, Vehoff:2008p1953}. 
A good review of observational results can be found for instance in 
\citet{Beuther:2007p1908}. 
Future generations of space telescopes and interferometric systems 
such as
the Herschel Space Observatory, 
the James Webb Space Telescope (JWST),
and the Atacama Large Millimeter Array (ALMA) 
will provide a deeper insight into the mechanisms of 
in-falling envelopes, 
bipolar outflows,
and massive accretion disks. 
This will definitely put tighter constraints on current theoretical models.

From the theoretical point of view,
assuming that the formation of high-mass stars is basically a scaled-up version of low-mass star formation, 
implies its own challenges:
Their rapid evolution, 
especially the shorter Kelvin-Helmholtz contraction timescale, 
leads to the interaction of the accreting flow of gas and
dust with the emitted radiation from the newborn
star 
\citep{Shu:1987p1616}. 
In the extreme case of a purely spherically symmetric collapse of a pre-stellar core (without angular momentum), 
the in-fall is potentially stopped by the growing radiation pressure entirely.
In the static limit, 
the equilibrium between radiative and gravitational forces at the
dust condensation radius is described by the Eddington limit 
$L_*/M_* = (4 \pi ~ G ~ c) / \kappa_*$,
where $L_*$, $M_*$, and $\kappa_*$ 
denote
the stellar luminosity,
mass, 
and the dust opacity regarding the stellar irradiation spectrum, 
respectively.
The underlying one-dimensional collapse problem is extensively studied in analytical studies
\citep{Penston:1969p2315, Shu:1977p2348, Hunter:1977p2352}
as well as in numerical simulations 
\citep{Penston:1969p2320, Hunter:1977p2352, Whitworth:1985p2244, Foster:1993p2412}, 
either in hydrodynamics or as evolution of self-similar solutions. 
Also stellar radiative feedback has been considered in one-dimensional studies 
\citep{Larson:1971p1210, Kahn:1974p1200, Yorke:1977p1358, Wolfire:1987p539}. 
However star formation is rarely a perfectly spherically symmetric process.

As stated by \citet{Nakano:1989p1267}, 
massive star formation is at least a two-dimensional non-spherically symmetric problem, 
including initial angular momentum to force the formation of a circumstellar disk and polar cavities.
Both will potentially help to overcome the Eddington barrier.
On small scales, 
\citet{Turner:2007p1126} show that the so-called photon bubble instability can
significantly enhance accretion and avoid the radiation pressure problem.
On large scales,
the formation of bipolar cavities possibly reduces the radiative
force onto the accretion flow by the so-called flashlight effect
\citep{Yorke:2002p1, Krumholz:2005p867}. 
Although
\citet{Krumholz:2005p24, Krumholz:2007p1380, Krumholz:2009p10975}
claim that the Eddington barrier may be broken via a Rayleigh-Taylor instability,
it is unclear if this instability is the most important one 
and 
how this instability will change in case of realistic frequency dependent radiative feedback instead of the used gray
(i.e.~frequency averaged) 
Flux Limited Diffusion 
(hereafter called FLD) 
approximation.

The important role of frequency dependent radiation transport in the formation of massive stars was already shown in 
\citet{Yorke:2002p1}, 
but due to the huge computational overhead of the frequency dependent radiation transfer it was
neither possible to study a huge amount of different initial conditions 
(to scan the parameter space), 
nor to perform high-resolution simulations of the accretion process.

Regarding these issues,
we demonstrate here the usability of our newly developed fast three-dimensional frequency dependent approximate
radiation transport module for numerical hydrodynamics.

The most accurate description of 
the physics proceeding during the collapse of a pre-stellar core
would include a
frequency dependent radiation transport step after each hydrodynamic timestep
using a modern Monte-Carlo based or ray-tracing radiative transfer method. 
The CPU time needed to solve one single hydrodynamic timestep is generally orders of magnitude lower 
than the time for a frequency dependent radiation transport step, 
especially in complex geometries.
Thus, 
this approach is not feasible with current computing technology for a large number of grid cells in more
than one dimension due to the large amount of computational time needed for each
radiation transport step. 
A more desired radiation hydrodynamics scheme should roughly spend 
the same CPU time on the radiation physics 
as on the hydrodynamics part. 
Sensible approximations are thus necessary to speed up the radiative transfer in such hydrodynamics studies.

In hydrostatic disk atmosphere computations,
\citet{Murray:1994p9750} 
introduced a splitting of a two-dimensional radiation field into an irradiated and a diffuse part.
Already 
\citet{Wolfire:1986p562, Wolfire:1987p539}
used such a splitting in one dimension to study the radiation feedback of massive stars, 
which has been shown to be a valid approach in one-dimensional hydrodynamics simulations by 
\citet{Edgar:2003p6}.
In the present paper,
we expand the method to higher dimensions and show its validity in an axially symmetric setup, 
consisting of a central star, 
a flared circumstellar disk,
and an envelope. 
In fact, 
we found that
it is necessary to consider 
the frequency dependence of the stellar irradiation feedback 
to reconstruct a reasonable approximation to the spatial temperature distribution.

A first approach for fast two-dimensional axially symmetric radiative transfer is,
for example,
the gray diffusion approximation studied by
\cite{Tscharnuter:1993p2109}, 
which is only applicable in the optically thick limit. 
Ray-tracing based methods
\citep[like in][]{Efstathiou:1990p2140} 
show high accuracy, 
but as already mentioned also require much CPU time, 
which yields low efficiency in combination with hydrodynamics solvers. 
Another common approach for the description of radiative processes in circumstellar disks is the gray FLD approximation
\citep{Kley:1989p2162, Bodenheimer:1990p2169, Klahr:1999p963, Klahr:2006p962}. 
It provides a fast method to determine the temperature evolution 
in the optically thick (diffusion) 
as well as in the optically thin (free-floating) limit, 
but shows stronger deviations in transition regions,
demonstrated e.g.~in 
\citet{Boley:2007p2959}.
In the case of gray FLD, 
which is still the default technique in modern radiation hydrodynamics codes,
this method of course suffers strongly from the lack of frequency dependence, 
when compared to the accuracy of modern Monte-Carlo based codes.

In the following, 
we present our results in combining the advantage of the gray FLD approximation (speed) 
with the accuracy of frequency dependent ray-tracing. 
The approximation described here results in a 
large reduction in computing time
compared to Monte-Carlo based radiative transfer.
This allows us
to implement this particular code in the framework of 
three-dimensional (magneto-) hydrodynamics simulations 
of circumstellar disks and in-falling envelopes on a parallel decomposed (spherical) grid. 
In particular,
we implemented our radiation transport module into 
the three-dimensional high-order Godunov magneto-hydrodynamics code Pluto
\citep{Mignone:2007p3421}. 
This tool allows us to perform high-resolution studies of accretion flows around a forming massive star.

The basic equations of the approximate radiation transport method 
and the numerical details 
are described in the following 
Sect.~\ref{sect:theory}. 
The test of the accuracy of the proposed radiation transport scheme 
in the setup of a currently standard radiation benchmark test
\citep{Pascucci:2004p39}
is presented in 
Sect.~\ref{sect:pascucci}. 
The parallel scaling behavior of our implementation is discussed in 
Sect.~\ref{sect:performance}. 
In 
Sect.~\ref{sect:radiativeshock},
we show the results of two standard radiation hydrodynamics shock tests 
performed with the FLD part of our implemented radiation transport method 
and the open source magneto-hydrodynamics code Pluto 
\citep{Mignone:2007p3421}. 
Conclusions as well as further studies planned are described in 
Sect.~\ref{sect:conclusions}.

\section{Theory and numerics of the approximate radiation transport scheme}
\label{sect:theory}
In the following, 
we recapitulate the general ideas, 
the basic equations,
and methods of the frequency dependent approximate radiation transport scheme. 
This section should allow the reader
to follow our motivation for the newly implemented three-dimensional radiative transfer module.
Every implemented formula and numerical detail is given.

The general idea of the method is to split the radiation transport into two components
\citep{Wolfire:1986p562, Murray:1994p9750, Edgar:2003p6}. 
Stellar radiative forces will mostly act on
the surrounding gas and dust mixture in the first transition region from optically thin 
(e.g.~where the dust is evaporated) 
to optically thick 
(e.g.~a massive accretion disk). 
This is exactly the region, where the FLD approximation is 
incorrect
\citep[e.g.][]{Boley:2007p2959}. 
To avoid this disadvantage of the FLD approximation, 
we first calculate the stellar radiative flux through the environment 
including its absorption in a corresponding first order ray-tracing routine.
First order means that
the spatial distribution of the radiative flux from stellar irradiation is calculated
according to its frequency dependent optical depth, 
but re-emission of photons is shifted to a gray FLD solver.
Sources for the thermal dust emission are 
the absorbed energy from the prior stellar irradiation step 
and potentially additional heating from the hydrodynamics,
e.g.~due to compression of the gas, 
accretion luminosity of sink cells, 
or viscous heating. 
In other words,
this means that instead of solving the whole radiation transfer problem either in the FLD approximation
or with a ray-tracing technique, 
we just extract the first (most important) absorption event of the stellar irradiation from the FLD solver
and calculate the appropriate flux in an accurate ray-tracing manner. 
This splitting method allows us to consider the frequency dependence of the stellar spectrum in a very
efficient manner.

In the first 
Subsect.~\ref{sect:fld}, 
we recapitulate
the FLD equation for thermal dust emission.
In the following two 
Subsects.~\ref{sect:irradiation} and \ref{sect:freqdepirradiation},
we explain how this FLD solver can simply be combined with a first order ray-tracing routine 
(either gray or frequency dependent respectively) 
to include irradiation feedback from a single central object. 
Due to the fact that these rays are aligned with the radial coordinate axis in spherical coordinates, 
this kind of coordinate system is highly favored, 
but not required, 
to solve the ray-tracing step. 
In the last
Subsect.~\ref{sect:gmres}, 
we comment on the so-called generalized minimal residual method (GMRES), 
our default implicit solver algorithm for the FLD equation.

\subsection{Flux Limited Diffusion}
\label{sect:fld}
The thermal dust emission is solved in the FLD approximation 
based on a diffusion equation for the thermal radiation energy $E_\mathrm{R}$. 
Within a given spatial density $\rho(\vec{x})$ and temperature $T(\vec{x})$ distribution, 
we start from the time evolution of the internal energy density $E_\mathrm{int}$ and thermal radiation energy
density $E_\mathrm{R}$:
\begin{eqnarray}
\label{eq:1}
\partial_t E_\mathrm{int} + \vec{\nabla} \cdot \left(E_\mathrm{int} ~ \vec{u} \right) 
&=& - P ~ \vec{\nabla} \cdot \vec{u} + \Lambda\\
\label{eq:2}
\partial_t E_\mathrm{R} + \vec{\nabla} \cdot \left(E_\mathrm{R} ~ \vec{u} \right) 
&=& - \vec{\nabla} \cdot \vec{F} - \Lambda
\end{eqnarray}
with the corresponding 
thermal pressure $P$, 
dynamical velocity $\vec{u}$, 
and flux of radiation energy density $\vec{F}$. 
The radiative heating and cooling of the gas is covered in 
$\Lambda = \rho ~ c ~ \kappa_\mathrm{R} ~ \left(a ~ T^4 - E_\mathrm{R} \right)$, 
where $c$ is the speed of light, 
$\kappa_\mathrm{R}$ the Rosseland mean opacity,
and $a$ the radiation constant.
Gas and dust temperatures are assumed to be the same 
$\left(T_\mathrm{gas} = T_\mathrm{dust} = T\right)$.

We add up the Eqs.~\eqref{eq:1} and \eqref{eq:2} and solve the transport term
$\vec{\nabla} \left(E_\mathrm{R} ~ \vec{u} \right)$ 
separately, if necessary, in the dynamical problem.
If we consider the transport of the internal energy 
$\vec{\nabla} \cdot \left(E_\mathrm{int} ~ \vec{u} \right)$ 
already during the corresponding hydrodynamics step (operator splitting), 
the remaining terms yield
\begin{equation}
\label{eq:3}
\partial_t \left(E_\mathrm{int} + E_\mathrm{R}\right) = - \vec{\nabla} \cdot \vec{F} + Q^+,
\end{equation}
where the source term 
$Q^+ =  - P ~ \vec{\nabla} \cdot \vec{u} + \ldots$ 
depends on the physics included and can additionally contain additional source terms such as
accretion luminosity from sink cells or viscous heating.

In the following, 
we use the assumption that the gas and radiation temperature are in equilibrium 
(also called one-temperature radiation transport), 
which is e.g.~a widely used approach in radiation hydrodynamics simulations of circumstellar disks. 
In a purely FLD radiation transport method, 
this assumption is justified for optically thick regions (e.g.~deeply inside an accretion disk). 
The usage of our split radiation
scheme guarantees also the correct gas temperature in regions of dominating stellar irradiation
(like an optically thin envelope or disk atmosphere). 
In residual regions, 
shielded from the irradiation behind the optically thick circumstellar disk, 
the gas and radiation temperature will be the same
in systems where the flow time is long compared to the
gas-radiation equilibrium time only. 
The flow time is $t_\mathrm{flow} \approx l / u$, 
where $l$ and $u$ are the characteristic length and velocity scales of the problem. 
The gas-energy equilibrium time is of the order of $t_\mathrm{eq} \approx (\kappa ~ \rho ~ c)^{-1}$, 
since the energy exchange rate is 
$\dot{E}_\mathrm{R} \approx \kappa ~ \rho ~ c ~ E_\mathrm{R}$. 
Thus, 
the requirement that $t_\mathrm{flow} \gg t_\mathrm{eq}$ reduces to the requirement that $\tau \gg u / c$, 
where $\tau \approx \kappa ~ \rho ~ l$ is the optical depth of the system. 
This is the regime that \citet{Mihalas:1984p9889} identifies as static diffusion. 
Thus,
the scheme is viable in the static diffusion limit. 
This is consistent with our test results for the radiative shock problem presented in Sect.~\ref{sect:radiativeshock}.
In this special case of a strong shock in a 
shielded, 
optically thin region,
the one-temperature approach yields the correct temperature of the shocked gas, 
but results in a steeper temperature gradient than a two-temperature radiation transport method in the
upstream direction of the radiation flux.
Consideration of such small
effects at outer regions of the domain goes beyond the scope of this research project, 
which focuses on the details of stellar radiative feedback on the accretion flow onto a massive star. 
Additionally,
\citet{Krumholz:2007p855}
show that massive star formation is generally in the static diffusion limit.
In the end, 
we will benefit from the speedup of the radiation transport solver by cutting the numbers of unknown variables in half. 
Moreover, 
the stiffness of the set of equations is much less if the above local equilibrium assumption is made.

Expressing both energies on the left hand side of Eq.~\eqref{eq:3} in terms of temperature allows us to
derive a relation between the time derivatives of the energies. 
The radiation energy density in absence of irradiation 
(see following Subsect.~\ref{sect:irradiation} for the case including irradiation) 
is given by 
\begin{equation}
\label{eq:4}
E_\mathrm{R} = a ~ T^4,
\end{equation}
This expression plus the internal energy density $E_\mathrm{int} = c_\mathrm{V} ~ \rho ~ T$ 
with the specific heat capacity $c_\mathrm{V}$ yield
\begin{equation}
\label{eq:5}
\partial_t E_\mathrm{int} 
= c_\mathrm{V} ~ \rho ~ \partial_t T 
= \frac{c_\mathrm{V} ~ \rho}{4 ~ a ~ T^3} \partial_t E_\mathrm{R}.
\end{equation}
With this relation, 
the Eq.~\eqref{eq:3} reduces to a standard diffusion equation 
\begin{equation}
\label{eq:6}
\partial_t E_\mathrm{R} =  f_\mathrm{c} \left( - \vec{\nabla} \cdot \vec{F} + Q^+ \right) 
\end{equation}
with $f_\mathrm{c} = \left(c_\mathrm{V} ~ \rho / \left(4 ~ a ~ T^3 \right) + 1\right)^{-1}$, 
depending only on the ratio of internal to radiation energy. 
The flux $\vec{F}$ of radiation energy density is determined in the FLD approximation via
\begin{equation}
\label{eq:7}
\vec{F} = - D ~ \vec{\nabla} E_\mathrm{R}
\end{equation}
with the diffusion constant $D = \lambda ~ c / \left(\kappa_\mathrm{R} ~ \rho \right)$.
The flux limiter $\lambda$ is chosen according to
\citet{Levermore:1981p57}.
Scattering is neglected.
In the most extreme limits the flux becomes either
$\vec{F} = - c ~ E_\mathrm{R} ~ \vec{\nabla} E_\mathrm{R} / |\vec{\nabla} E_\mathrm{R}|$
for highly optically thin regions (free-streaming limit) or
$\vec{F} = - c ~ \vec{\nabla} E_\mathrm{R} / \left(3 ~ \kappa_\mathrm{R} ~ \rho \right)$
for highly optically thick regions (diffusion limit).

Summing up, 
we can describe the thermal/diffuse radiative processes via:
\begin{equation}
\label{eq:8}
\partial_t E_\mathrm{R} = - f_\mathrm{c} \left( \vec{\nabla} \cdot \vec{F} - Q^+ \right)
\end{equation}
with
\begin{eqnarray}
\label{eq:9}
f_\mathrm{c} &=& \left(c_\mathrm{V} ~ \rho / \left(4 ~ a ~ T^3 \right) + 1 \right)^{-1} \\
\label{eq:10}
\vec{F} &=& - D ~ \vec{\nabla} E_\mathrm{R} \\
\label{eq:11}
D &=& \lambda ~ c / \left(\kappa_\mathrm{R} ~ \rho \right) \\
\label{eq:12}
\lambda &=& \left(2+R \right) / \left(6+3R+R^2 \right) \\
\label{eq:13}
R &=& |\vec{\nabla} E_\mathrm{R}| / \left(\kappa_\mathrm{R} ~ \rho ~ E_\mathrm{R} \right) \\
\label{eq:14}
Q^{+} &=&  - P ~ \vec{\nabla} \cdot \vec{u} + \mbox{ additional source terms}
\end{eqnarray}

\subsection{Irradiation}
\label{sect:irradiation}
When including stellar irradiation, 
some of the equations have to be modified.
The irradiation from a central object is treated as an additional flux $\vec{F}_*$, 
released along rays in the radially outward direction. 
$E_\mathrm{R}$ is the energy density of the radiation emitted by the circumstellar material, 
and does not include the stellar irradiation. 
The additional radiation power
$- \vec{\nabla} \cdot \vec{F}_*$
from this irradiation is therefore added to the diffusion Eq.~\eqref{eq:8} as a source term
\begin{equation}
\label{eq:15}
\partial_t E_\mathrm{R} = - f_\mathrm{c} \left( \vec{\nabla} \cdot \vec{F} + \vec{\nabla} \cdot \vec{F}_* - Q^+ \right)
\end{equation}
and has to be added to the left hand side of Eq.~\eqref{eq:4}, 
which is used to calculate 
either the thermal radiation density 
or the corresponding dust temperature $T$ 
under the assumption that the dust is in equilibrium with the combined stellar and diffuse radiation field:
\begin{equation}
\label{eq:16}
a ~ T^4 = E_\mathrm{R} + \frac{\kappa_\mathrm{P}(T_*)}{\kappa_\mathrm{P}(T)} \frac{|\vec{F}_*|}{c}.
\end{equation}
Here $\kappa_\mathrm{P}(T)$ represents the Planck mean opacity to a given temperature $T$ 
and $T_*$ is the star's effective temperature. 

After solving the diffusion Eq.~\eqref{eq:15} implicitly 
(with the radiation energy $E_\mathrm{R}\left(\vec{x}\right)$ as the solution vector),
the corresponding dust temperature is calculated using Eq.~\eqref{eq:16}.
Since the opacity $\kappa_\mathrm{P}\left(T\right)$ depends on the temperature the right hand side of Eq.~\eqref{eq:16}
also depends on the solution of $T$. 
This makes an iterative procedure based on the Newton-Raphson method necessary to find the solution.
The ratio of opacities 
$\kappa_\mathrm{P}\left(T_*\right) / \kappa_\mathrm{P}\left(T\right)$ 
corresponds to the relation of emission from and absorption by dust.

The stellar radiative flux as a function of distance $r$ from the central object is calculated by
\begin{equation}
\label{eq:17}
\vec{F}_*\left(r\right) = \vec{F}_*\left(R_*\right) ~ \left(\frac{R_*}{r}\right)^2
\exp\left(-\tau\left(r\right)\right) 
\end{equation}
The (boundary) flux at the stellar surface $\vec{F}_*\left(R_*\right)$ can be calculated from the Stefan-Boltzmann law
\begin{equation}
\label{eq:18}
|\vec{F}_*\left(R_*\right)| = \sigma_{SB} ~ T_*^4
\end{equation}
with the Stefan-Boltzmann constant
$\sigma_{SB} = 0.25 ~ a ~ c$.
The optical depth $\tau$ is, 
in the case of gray irradiation 
(see following Subsect.~\ref{sect:freqdepirradiation} for specific changes due to frequency dependent irradiation), 
calculated along radial rays through the spherical grid via
\begin{equation}
\label{eq:19}
\tau\left(r\right) = \int_{R_*}^r \kappa_\mathrm{P}\left(T_*\right) ~ \rho\left(r\right) ~ dr.
\end{equation}
In the case of the radiation benchmark test by
\cite{Pascucci:2004p39},
the optical depth between the stellar surface and the inner disk 
(the inner boundary of the spherical computational domain) 
is assumed to be negligible.

A flow chart of the radiation module described so far is shown for a static problem 
such as the benchmark test of Sect.~\ref{sect:pascucci}
in Fig.~\ref{flow-chart}.

\setlength{\unitlength}{1cm}
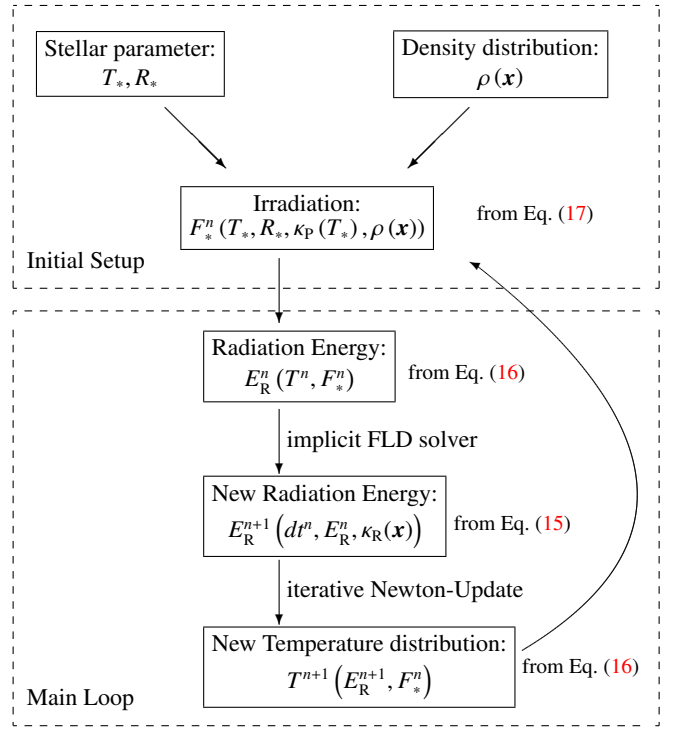
\begin{figure}[h]
\begin{picture}(8, 9.5)
\fboxsep1mm

\put(0.3, 8.5){\fbox{\shortstack{Stellar parameter: \\ $T_*, R_*$}}}
\put(5.0, 8.5){\fbox{\shortstack{Density distribution: \\ $\rho\left(\vec{x}\right)$}}}
\put(2.2, 6.5){\fbox{\shortstack{Irradiation: \\ $F_*^n\left(T_*, R_*, \kappa_\mathrm{P}\left(T_*\right),
\rho\left(\vec{x}\right)\right)$}}} 
\put(2.5, 4.5){\fbox{\shortstack{Radiation Energy: \\ $E_\mathrm{R}^n\left(T^n,
F_*^n\right)$}}} \put(2.5, 2.5){\fbox{\shortstack{New Radiation Energy: \\ 
$E_\mathrm{R}^{n+1}\left(dt^n, E_\mathrm{R}^n, \kappa_\mathrm{R}(\vec{x})\right)$}}}
\put(2.5, 0.5){\fbox{\shortstack{New Temperature distribution: \\ $T^{n+1}\left(E_\mathrm{R}^{n+1}, F_*^n\right)$}}}

\put(2.0, 8.2){\vector(1, -1){0.8}}
\put(6.0, 8.2){\vector(-1, -1){0.8}}
\put(3.5, 6.2){\vector(0, -1){0.86}}
\put(3.5, 4.2){\vector(0, -1){0.86}}
\put(3.5, 2.2){\vector(0, -1){0.86}}

\bezier{600}(6.7, 1.0)(10.0, 3.0)(6.1, 6.1)
\put(6.1, 6.1){\vector(-2, 1){0.1}}

\put(3.6, 3.7){implicit FLD solver}
\put(3.6, 1.7){iterative Newton-Update}

\put(0.0, 5.8){\dashbox{0.1}(8.5, 3.75)[bl]{\hspace{1mm} Initial Setup \vspace{2mm}}}

\put(0.0, 0.0){\dashbox{0.1}(8.5, 5.5)[bl]{\hspace{1mm} Main Loop \vspace{2mm}}}

\put(6.1, 6.7){\tiny from Eq.~\eqref{eq:17}}
\put(5.2, 4.6){\tiny from Eq.~\eqref{eq:16}}
\put(5.8, 2.6){\tiny from Eq.~\eqref{eq:15}}
\put(6.7, 0.7){\tiny from Eq.~\eqref{eq:16}}

\end{picture}
\caption{
Schematic flow chart of the radiation module.
Exponents declare the iteration (timestep) number $n$.
The actual timestep corresponds to $dt^n = t^{n} - t^{n-1}$.}
\label{flow-chart}
\end{figure}

\subsection{Frequency dependent irradiation}
\label{sect:freqdepirradiation}
\begin{figure}[t]
\begin{center}
\includegraphics[width=0.48\textwidth]{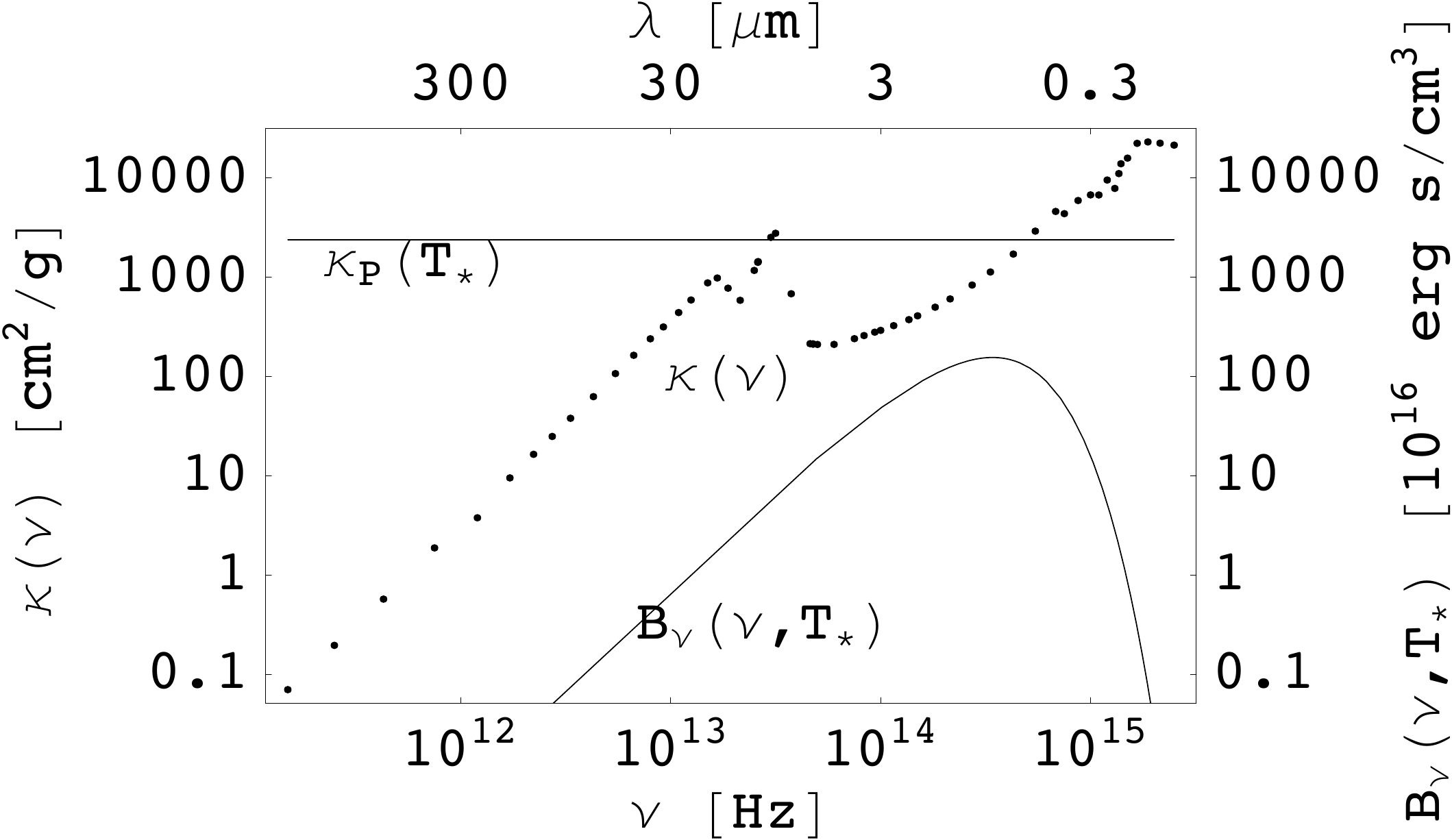}
\caption{
Regarding the frequency dependence of stellar irradiation feedback.
Frequency dependent opacities $\kappa\left(\nu\right)$, 
Planck mean opacities $\kappa_\mathrm{P}\left(T_*\right)$, 
and stellar black body spectrum $B_\nu\left(\nu,T_*\right)$ as function of the frequency $\nu$.
}
\label{Opacities}
\end{center}
\end{figure}

Taking into account the frequency dependence of the stellar flux, 
we consider a fixed number of frequency bins,
characterized by their mid-frequency $\nu_i$, instead of the gray irradiation. 
For the radiation benchmark test we use the opacity tables of 
\citet{Draine:1984p594}, 
including 61 frequency bins,
see Fig.~\ref{Opacities}. 
Each dot in Fig.~\ref{Opacities} represents the mid-frequency of the corresponding frequency bin.
Including the frequency dependence,
we have to replace the Planck mean opacity $\kappa_\mathrm{P}\left(T_*\right)$ 
in Eqs.~\eqref{eq:16} and \eqref{eq:19} 
with the frequency dependent opacities $\kappa\left(\nu\right)$ 
and the stellar flux $\vec{F}_*$ in
Eqs.~\eqref{eq:15} and \eqref{eq:16} 
with the corresponding sum over the frequency dependent fluxes $\vec{F}_*\left(\nu\right)$, 
each calculated via Eq.~\eqref{eq:17}.

The boundary condition, 
given by Eq.~\eqref{eq:18}, 
is now determined for each frequency bin by the integral over the corresponding part of the stellar black body Planck
function $B_\nu\left(\nu, T_*\right)$ in the frequency dependent interval:
\begin{equation}
\label{eq:20}
|\vec{F}_*\left(R_*, \nu_i\right)| = \frac{c}{4} \int\limits_{\left(\nu_{i-1}+\nu_i\right)/2}^{\left(\nu_i +
\nu_{i+1}\right)/2} B_\nu\left(\nu, T_*\right) ~ d\nu
\end{equation}

\subsection{The generalized minimal residual method}
\label{sect:gmres}
The FLD Eqs.~\eqref{eq:8} or \eqref{eq:15} are solved in our newly implemented approximate radiation transport module for
hydrodynamics in an implicit fashion via a generalized minimal residual (GMRES) solver. 
The GMRES solver is a Krylov subspace (KSP) method 
for solving a system of linear equations  $A \vec{x} = \vec{b}$ 
via an approximate inversion of the large but sparse matrix $A$ 
and is an advancement of the minimal residual (MinRes) method. 
The GMRES method was developed in 1986 and is described in \citet{Saad:1986p2176}. 
This method is much better than the conjugate gradient (CG) 
or successive over-relaxation (SOR) method, 
and at least as good as the well-known improved stabilized Bi-Conjugate Gradient (BiCGstab) method used, 
e.g., in
\citet{Yorke:2002p1}.

The general idea of minimal residual solvers based on the KSP method is the following:
The $i^\mathrm{th}$ KSP is defined as 
$K_i = span\{\vec{b}, A \vec{b}, A^2 \vec{b}, ..., A^{i-1} \vec{b}\}$. 
In each solver iteration $i$ 
the GMRES method increments the subspace $K_i$ used with an additional basis vector
$A^{i-1} \vec{b}$ 
and approximates the solution of the system of linear equations by the vector $\vec{x}_i$ 
which minimizes the norm of the residual $r = |A \vec{x}_i - \vec{b}|$. 
This method converges monotonically 
and theoretically reaches the exact solution after performing as many iterations as the column number of the matrix $A$ 
(which equals the number of grid cells). 
Of course, 
in practice the iteration is already stopped after reaching a specified relative or absolute tolerance of the residual, 
which normally takes only a small number of iterations.
The computation of each iteration grows like $O\left(i^2\right)$.
In the current implementation, 
we use the so-called 'GMRES restarted' by default.
GMRES restarted never performs all the iterations to reach the exact solution.
After a priorily fixed number of internal iterations $n$, 
the solver starts a second time in the first
subspace $K_1$ but with the last approximate solution $\vec{x_n}$. 
Due to the growing computational effort with $O\left(i^2\right)$ this approach generally results in a speedup of the
computation.

The radiation transport module is parallelized for distributed memory machines, 
using the message passing interface (MPI) language. 
The results of a detailed parallel performance test of the whole radiation transport module, 
including this GMRES solver is presented in Sect.~\ref{sect:performance}.

\section{Frequency dependent test of the approximate radiation transport}
\label{sect:pascucci}
The approximate radiation transport introduced in the previous section can now be tested for realistic dust
opacities in a standard benchmark test for irradiated circumstellar disk models.
The setup of the following comparison is adopted from
\citet{Pascucci:2004p39}
and includes 
a central solar-type star, 
an irradiated circumstellar flared disk 
and an envelope.
We have to choose a low-mass central star, 
because no benchmark for high-mass stars was performed so far.
However,
the tests should not depend on the actual size and luminosity of the central star.
For comparison, 
we choose a standard full frequency dependent Monte-Carlo based radiation transport code.
The comparison is done for two different 
(low and high) 
optical depths taken from the original radiation benchmark test.
To test each of the components 
(gray and frequency dependent irradiation as well as Flux Limited Diffusion) 
of the proposed approximate radiative transfer separately, 
we perform several test runs with and without
the different physical processes 
(absorption and re-emission) 
with both 
the Monte-Carlo based 
(see table~\ref{MC-runs}) 
and the approximate 
(see table~\ref{run-table}) 
radiative transfer code.

\subsection{Setup}
\label{sect:setup}
\subsubsection{Physical setup of the star, the disk, and the envelope}
\noindent
The stellar parameters are solar-like:
The effective temperature of the star is 5800 K $\left(T_* = 5800 \mbox{ K}\right)$
and the stellar radius is fixed to 1 solar radius $\left(R_* = 1 \mbox{ R}_\odot\right)$.
The disk ranges from $r_\mathrm{min} = 1$ AU up to $r_\mathrm{max} = 1000$ AU.

\noindent
Although the numerical setup of the gas density is done in spherical coordinates, 
the analytic setup of the gas density, 
as described in the original benchmark test, 
is given in cylindrical coordinates:
\begin{equation}
\rho\left(r, z\right) = \rho_0 ~ f_1\left(r\right) ~ f_2\left(r, z\right)
\end{equation}
with the radially and vertically dependent functions
\begin{equation}
f_1\left(r\right) = \frac{r_\mathrm{d}}{r}
\end{equation}
and
\begin{equation}
f_2\left(r, z\right) = \exp \left(- \frac{\pi}{4} ~ \left( \frac{z}{h\left(r\right)}\right)^2 \right)
\end{equation}
making use of the following abbreviations
\begin{eqnarray}
h\left(r\right) &=& z_\mathrm{d} ~ \left( \frac{r}{r_\mathrm{d}} \right)^{1.125},\\
r_\mathrm{d} &=& \frac{r_\mathrm{max}}{2} = 500 \mbox{ AU} \mbox{, and}\\
z_\mathrm{d} &=& \frac{r_\mathrm{max}}{8} = 125 \mbox{ AU.}
\end{eqnarray}
The lowest density is limited to a relative factor of $10^{-100}$ compared to the highest density 
(at $r_\mathrm{min}$ in the midplane) 
to avoid divisions by zero 
(e.g.~in the calculation of the diffusion coefficients). 
The normalization $\rho_0$ of the density setup is chosen to define different optical depths 
$\tau_{550\mathrm{nm}}$
through the midplane of the corresponding circumstellar disk 
(at a visual wavelength of $550$ nm):
\newline

\renewcommand{\arraystretch}{1.4}
\begin{tabular}{| l | l | l |}
\hline
$\tau_{550\mathrm{nm}}$	& $\rho_0 ~ [\mbox{g} \mbox{ cm}^{-3}]$	& $M_\mathrm{tot}$ of gas $[\mbox{M}_\odot]$ \\
\hline \hline
$0.1$	                   & $8.321*10^{-21}$		& $1.1 * 10^{-5}$ \\
$100$	                   & $8.321*10^{-18}$		& $1.1 * 10^{-2}$ \\
\hline
\end{tabular}
\renewcommand{\arraystretch}{1}
\newline

\noindent
The opacity tables used are the same as in the original benchmark test
\citep{Pascucci:2004p39}
taken from
\citet{Draine:1984p594}.
They are displayed in Fig.~\ref{Opacities}.

\subsubsection{Numerical setup of the approximate radiation transfer code}
The runs of the approximate radiative transfer code are performed on a 
radially stretched, 
polar uniform,
spherical, 
two-dimensional grid. 
The grid consists of 60 cells in the radial direction 
times 61 cells in the polar direction 
(plus additional cells for the storage of boundary conditions). 
The polar range covers the full spatial setup of $180\degr$.

Stretching of the radial grid dimension by an
additional 10\% from one cell to the next is applied.
The implicit diffusion Eq.~\eqref{eq:15} is solved via the GMRES method 
(see Sect.~\ref{sect:gmres}) 
after parallel/global Block-Jacobian 
and serial/local ILU pre-conditioning 
in the framework of the version 2.3.3 of the open source parallel solver library PETSc 
(Portable, Extensible Toolkit for Scientific Computation). 
More detailed information about this solver library can be found in \citet{petsc-web-page, petsc-user-ref}.

The gradient of the radiation energy is zero at the inner radial and both polar boundaries
$\left( \vec{\nabla} E_\mathrm{R} = 0 \right)$, 
i.e.~radiative flux over these boundaries is prohibited. 
The radial outer boundary is defined as a constant Dirichlet boundary corresponding to 
$T_0 = 14.7 \mbox{ K}$ $\left( E_\mathrm{R} = a ~ T_0^4 \right)$.

The timestep used for the FLD solver is $10^4$ s.
The main iteration 
(circle of irradiation and FLD steps) 
is stopped when the relative change of the temperature in each cell is smaller than 0.01\%, 
leading to 3 main iterations in the ``purely absorption'' runs
and more than 600 main iterations in the ``irradiation plus FLD'' runs.

\subsubsection{Numerical setup of the Monte-Carlo based comparison code RADMC}
For comparison we use the Monte-Carlo based radiation transfer code RADMC described in 
\citet{Dullemond:2000p2185} and \citet{Dullemond:2004p2184}.
The general solver method of the code is based on
\citet{Bjorkman:2001p2187}.
The Monte-Carlo runs are performed on a 60 x 31 grid assuming symmetry to the disk midplane.
The grid is stretched in both directions.
One million photons are used.

Scattering can be handled by this code, but is simply switched off as it is neglected in the radiation transport
module described here. 
Scattering would increase the temperature in the irradiated parts 
(up to an optical depth of about unity) 
by about 2\% in the optically thin envelope 
up to a maximum of 19\% in the optically thick inner rim of the disks midplane due to higher extinction. 
The more effectively shielded outer regions of the disk would be about 10\% cooler. 
For more massive and luminous stars the effect of scattering would decrease due to stronger forward
scattering, which is included in our ray-tracing routine per definition.

\subsection{Configurations of runs performed}
\label{sect:configuration}
\begin{table*}[t]
\begin{center}
\begin{tabular}{| l || c | c | c |}
\hline
Run                            & MC0.1-full    & MC100-A               & MC100-full     \\
\hline\hline
Optical depth $\tau_{550\mathrm{nm}}$   & 0.1           & 100                   & 100            \\
RADMC - Configuration          & ``full''          & ``one-photon-limit''    & ``full''           \\
frequency dependence           & yes           & yes                   & yes            \\
\hline
Comparison Sect.               & \ref{sect:irradiation_thin}         & \ref{sect:irradiation_thick}                 &
\ref{sect:full_thick} \& \ref{sect:radiativeforce} \\
\hline
\end{tabular}
\end{center}
\caption{
Overview of performed Monte-Carlo runs for comparison.
\newline
The overview table of the comparison runs performed with the Monte-Carlo based code RADMC contains 
the corresponding optical depth of the test case 
and 
the used configuration of the Monte-Carlo code,
see referred comparison Sects.~for details on the ``one-photon-limit'' and ``full run''. 
All Monte-Carlo runs include frequency dependence. 
}
\label{MC-runs}
\end{table*}

\begin{table*}[t]
\begin{center}
\begin{tabular}{| l || c | c | c | c | c |}
\hline
Run                           & G0.1       & G100                     & F100                    & GD100                
& FD100\\ 
\hline\hline
Optical depth $\tau_{550\mathrm{nm}}$  & 0.1        & 100                      & 100                     &
100                      & 100\\
\hline
Comparison run                & MC0.1-full & MC100-A                  & MC100-A                 & MC100-full         
& MC100-full\\
\hline
Ray-tracing Irradiation       & gray       & gray                     & freq.~dep.              & gray                 
& freq.~dep.\\ 
Flux Limited Diffusion & no         & no                      & no                      & yes                      &
yes\\
\hline
Theory Sect.                  & \ref{sect:irradiation}        & \ref{sect:irradiation}                      &
\ref{sect:irradiation} \& \ref{sect:freqdepirradiation} & \ref{sect:fld} \& \ref{sect:irradiation} & \ref{sect:fld} -
\ref{sect:freqdepirradiation}\\ 
Comparison Sect.              & \ref{sect:irradiation_thin}      & \ref{sect:irradiation_thick}                   
& \ref{sect:irradiation_thick} & \ref{sect:full_thick} & \ref{sect:full_thick} \& \ref{sect:radiativeforce}\\
\hline
Deviation $\left(\Delta T\right)/T$ [\%] & $<$ +2.0   & +10.9 $\rightarrow$ -57.2 & +5.0 $\rightarrow$ -0.6 & +5.0
$\rightarrow$ -38.4 & +9.3 $\rightarrow$ -11.1\\
\hline
\end{tabular}
\end{center}
\caption{
Overview of runs using the proposed approximate radiation transport module.
\newline
The overview table of the runs discussed contains 
the corresponding optical depth of the test case and the Monte-Carlo run, 
which is used for comparison 
(see also table~\protect\ref{MC-runs} and referred theory and comparison Sects.).
Furthermore, 
the applied radiative modules 
(gray or frequency dependent absorption as well as possible diffusion) 
of our proposed approximate radiation transport method 
and the corresponding Sects., 
in which this modules and the final results are discussed, 
are given. 
The deviations in the resulting temperature profiles of the approximate radiation transport from the corresponding
Monte-Carlo comparison run are shown in the lower row. 
}
\label{run-table}
\end{table*}

The following comparison of the results is divided into three parts 
(each of the proposed components of the theoretical 
Sects.~\ref{sect:fld} - \ref{sect:freqdepirradiation} are tested independently): 
First, 
we study a pure absorption
scenario without diffusion in the optically thin and thick case. 
Afterwards we include diffusion effects.
Therefore, 
we perform three different runs of the Monte-Carlo based code:
a full run for the optically thin case $\tau_{550\mathrm{nm}} = 0.1$,
a full run for the optically thick case $\tau_{550\mathrm{nm}} = 100$,
and an additional run for the optically thick case $\tau_{550\mathrm{nm}} = 100$
with excluded re-emission of the photons 
(to achieve a pure absorption scenario for comparison with our first order ray-tracing routine),
hereafter called ``one-photon-limit''.
An overview of these three comparison runs is given in table~\ref{MC-runs}.

The resulting temperature distributions of these Monte-Carlo runs are compared with 
the results
of the approximate radiative transfer runs including different components of our module: 
We discuss five different configurations for 
the optically thin and thick setup including gray and frequency dependent irradiation plus potential diffusion.
An overview of the physics applied 
and the resulting deviations of these runs from the comparison data 
is given in table~\ref{run-table}. 
These results are discussed and illustrated in detail in the following subsections.

\subsection{Results}
\label{sect:results}
\subsubsection{Gray absorption in an optically thin disk}
\label{sect:irradiation_thin}
\begin{figure}[t]
\begin{center}
\includegraphics[width=0.48\textwidth]{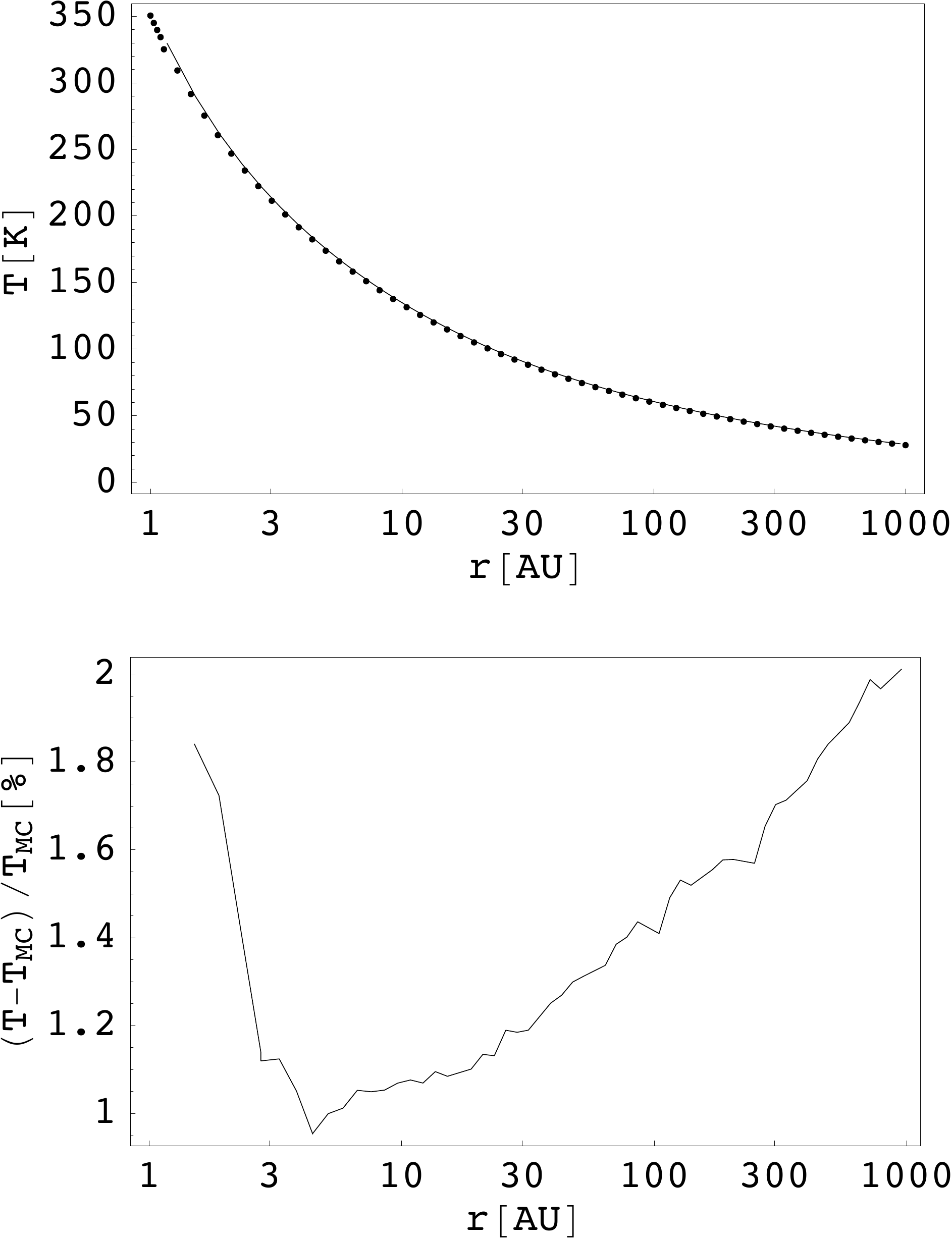}
\caption{
Radial cut through the temperature profile in the midplane in the most optically thin case $\tau_{550\mathrm{nm}} = 0.1$.
\newline
Upper panel: 
Radial temperature slope of the gray irradiation routine (solid line) and the Monte-Carlo based comparison code (dots).
\newline
Lower panel: Differences between the two codes in percent.
}
\label{Tau0001}
\end{center}
\end{figure}

In the most optically thin case $\tau_{550\mathrm{nm}} = 0.1$, 
diffusion effects should be negligible. 
Therefore, 
we can test the validity of the routines described in 
Sect.~\ref{sect:irradiation} 
without running the diffusion routine. 
Compared to the full Monte-Carlo simulation from RADMC 
also the deviation in the most ``difficult'' region of the midplane 
(due to having the highest absorption) 
stays below 2\% (see Fig.~\ref{Tau0001}).

In this optically thin limit, 
the diffusion effects are indeed negligible:
An additionally performed full run of the approximate radiation transfer module 
with frequency dependent irradiation plus Flux Limited Diffusion shows a variation in the radial temperature slope from
the pure gray irradiation run below 1\%.

\subsubsection{Gray and frequency dependent absorption in an optically thick disk}
\label{sect:irradiation_thick}
In the most optically thick case $\tau_{550\mathrm{nm}} = 100$, 
we test a pure absorption case to distinguish the deviations introduced by the FLD approximation in the full run from
the deviations introduced by the irradiation component of our module. 
\begin{figure}[t]
\begin{center}
\includegraphics[width=0.48\textwidth]{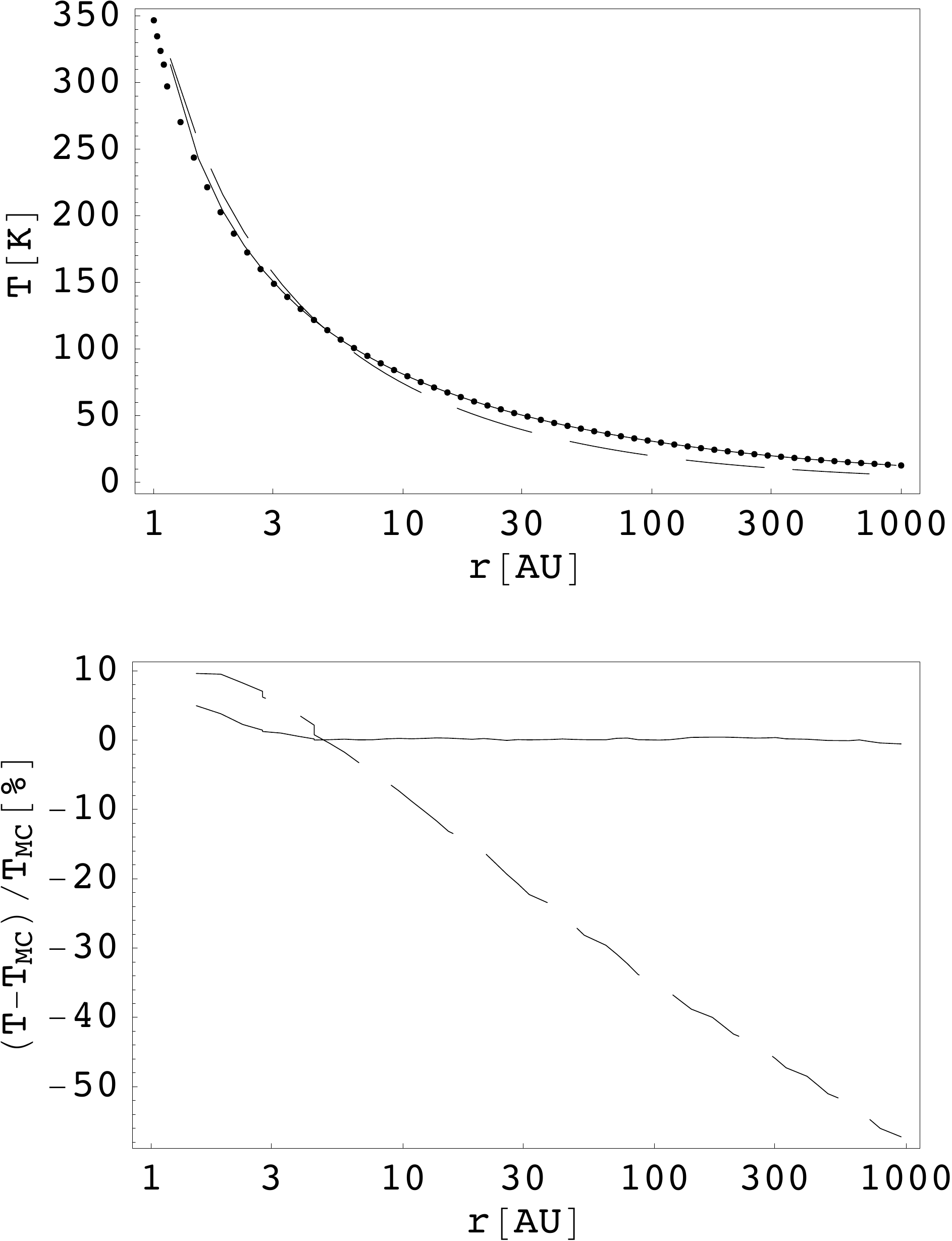}
\caption{
Radial cut through the temperature profile in the midplane in the most optically thick case 
$\tau_{550\mathrm{nm}} = 100$ without diffusion.
\newline
Upper panel: 
Radial temperature slope of gray irradiation (dashed line), 
frequency dependent irradiation (solid line),
and the Monte-Carlo routine in the "one-photon-limit" (dots).
\newline
Lower panel: 
Deviations of the gray (dashed line) 
and frequency dependent method (solid line) 
from the Monte-Carlo code in percent. 
}
\label{Tau1000ab}
\end{center}
\end{figure}
Therefore, 
we run the Monte-Carlo code only until every initial photon is absorbed, 
neglecting re-emission or scattering events. 
In this scenario,
we are able to probe the absorption routines of 
Sects.~\ref{sect:irradiation} and \ref{sect:freqdepirradiation} 
in detail 
and determine the improvement by considering the frequency dependence of the stellar irradiation. 
The resulting temperature 
profiles
through the midplane for
the case of gray and frequency dependent irradiation
as well as the corresponding deviations from the Monte-Carlo comparison run are shown in Fig.~\ref{Tau1000ab}.
Indeed, 
we found that in order to limit the deviation in the absorption part of the radiation module to less than 5\% 
it is essential to account for the frequency dependence of the stellar irradiation. 
That is due to the fact that the infrared part of the
stellar spectrum has a lower optical depth than the UV part.
Neglecting the frequency dependence results roughly in a 57\% cooler disk at large radii and a 10\% hotter inner rim
compared to the Monte-Carlo based radiative transfer code. 

\subsubsection{Gray and frequency dependent absorption plus diffusion in an optically thick disk
(the complete problem)}
\label{sect:full_thick}
In this and the following section, 
we show the comparison results between the complete approximate radiation transport
method with the combination of 
(gray or frequency dependent) 
irradiation and Flux Limited Diffusion and the
corresponding Monte-Carlo simulation in the most optically thick case $\tau_{550\mathrm{nm}} = 100$.

\begin{figure}[t]
\begin{center}
\includegraphics[width=0.48\textwidth]{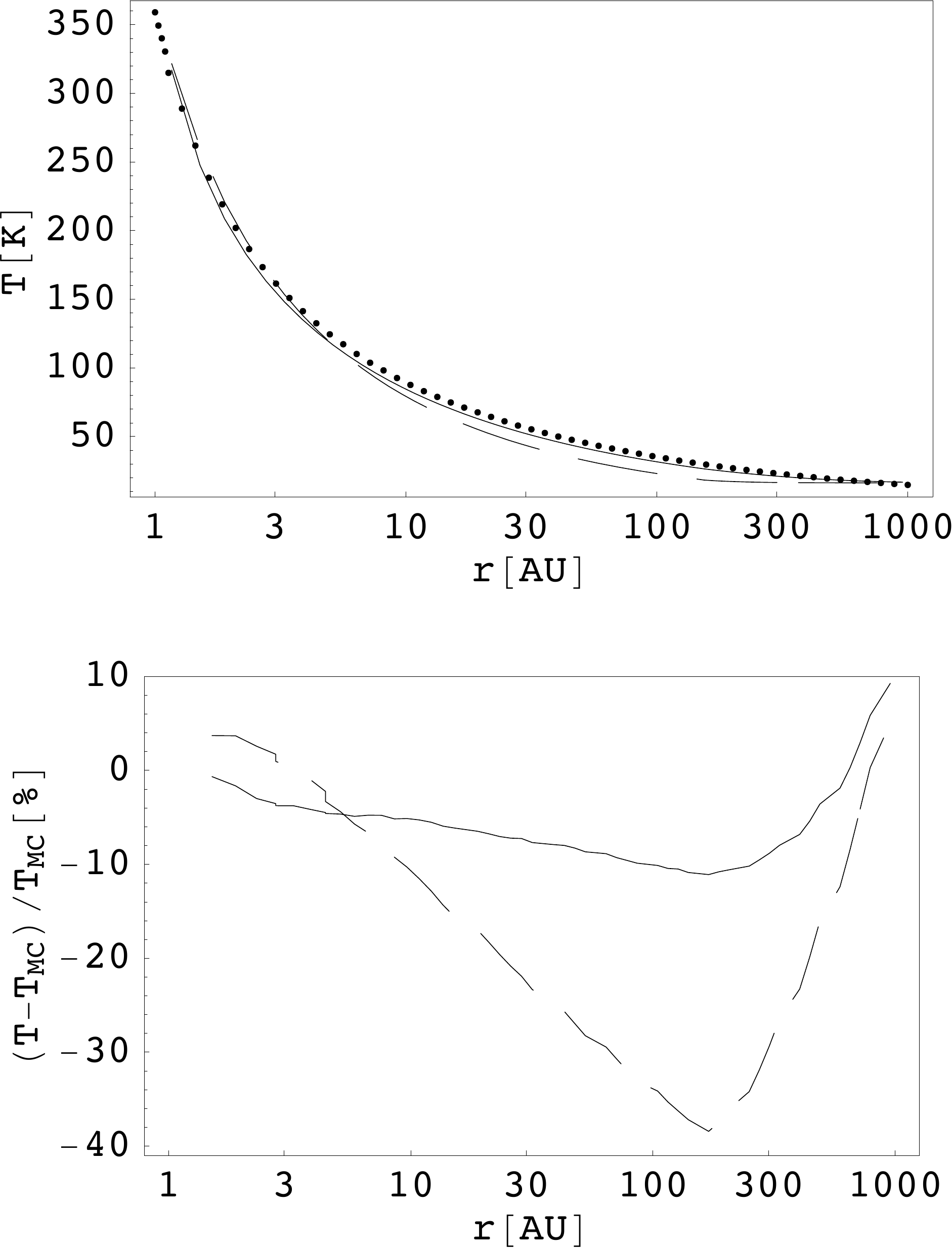}
\caption{
Radial cut through the temperature profile in the midplane in the most optically thick case 
$\tau_{550\mathrm{nm}} = 100$ including irradiation and Flux Limited Diffusion.
\newline
Upper panel: 
Radial temperature slope of gray irradiation plus Flux Limited Diffusion (dashed line), 
frequency dependent irradiation plus Flux Limited Diffusion (solid line),
and the corresponding Monte-Carlo routine (dots).
\newline
Lower panel: 
Deviations of the gray (dashed line) 
and frequency dependent run (solid line) 
from the Monte-Carlo code in percent.
}
\label{Tau1e+2_Irradiation+FLD_radial}
\end{center}
\end{figure}

\begin{figure}[t]
\begin{center}
\includegraphics[width=0.48\textwidth]{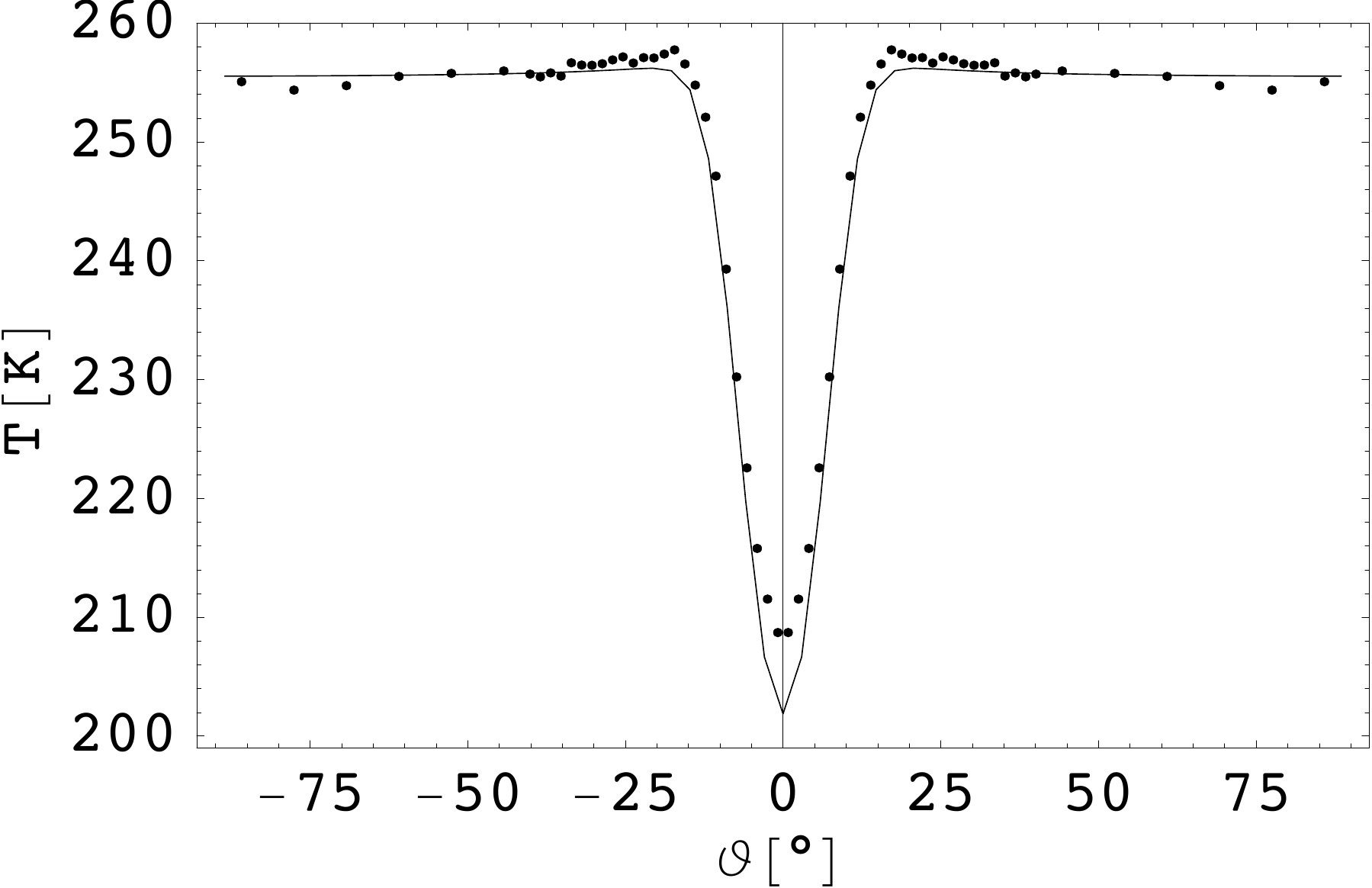}
\caption{
Polar cut through the temperature profile at $r=2$ AU of the frequency dependent irradiation plus Flux Limited Diffusion
run for the most optically thick case $\tau_{550\mathrm{nm}} = 100$.
The profile reproduces the turnover point at a polar angle of $\theta \approx 19\degr$ above the midplane from the
optically thin envelope to the optically thick disk region.
\newline
Solid line: Frequency dependent irradiation plus Flux Limited Diffusion.
\newline
Dots: Data from the corresponding Monte-Carlo comparison run.
\newline
The vertical axis covers only a small temperature range from 200 to 260~K to better visualize the small deviations.
}
\label{Tau1e+2_polar}
\end{center}
\end{figure}

Also in this case, 
we found that the frequency dependent irradiation is necessary to achieve
the accuracy needed for realistic radiative feedback in hydrodynamics studies 
(e.g.~in massive star formation or in irradiated accretion disks). 
Gray irradiation combined with the FLD approximation leads to deviations up to 38.4\% in
the resulting temperature slope 
(see Fig.~\ref{Tau1e+2_Irradiation+FLD_radial}). 
Including frequency dependent irradiation and FLD the radial temperature profiles
agree within 0.7\% at the inner boundary up to a maximum deviation of 11.1\% at roughly $r \approx 200 \mbox{ AU}$. 
The comparison between the complete radiation scheme and the corresponding Monte-Carlo run is shown in radial and 
polar temperature profiles in 
Figs.~\ref{Tau1e+2_Irradiation+FLD_radial} and \ref{Tau1e+2_polar} respectively. 
The turnover point 
(at a polar angle from the midplane of $\theta \approx 19\degr$) 
from the optically thin envelope to the optically thick disk region is reproduced very well.

Further remarks and discussion of these results are given in the last Subsect.~\ref{sect:remarks} at the end of the
comparison section.

\subsubsection{Radiative force}
\label{sect:radiativeforce}
Stellar radiative feedback on the dynamics of the environment plays a crucial role in the formation of massive stars.
The heating will probably prevent further fragmentation of the cloud by enhancing the Jeans mass
\citep[e.g.][]{Krumholz:2007p1380}.
Furthermore, 
the dusty environment feels the radiative force when absorbing the radiation due to momentum conservation,
which potentially stops the accretion process for highly luminous massive stars. 
Therefore,
it is necessary to compute the correct radiative flux (and its derivative, the radiative force) in addition to the
temperature distribution in such simulations.

\begin{figure}[t]
\begin{center}
\includegraphics[width=0.48\textwidth]{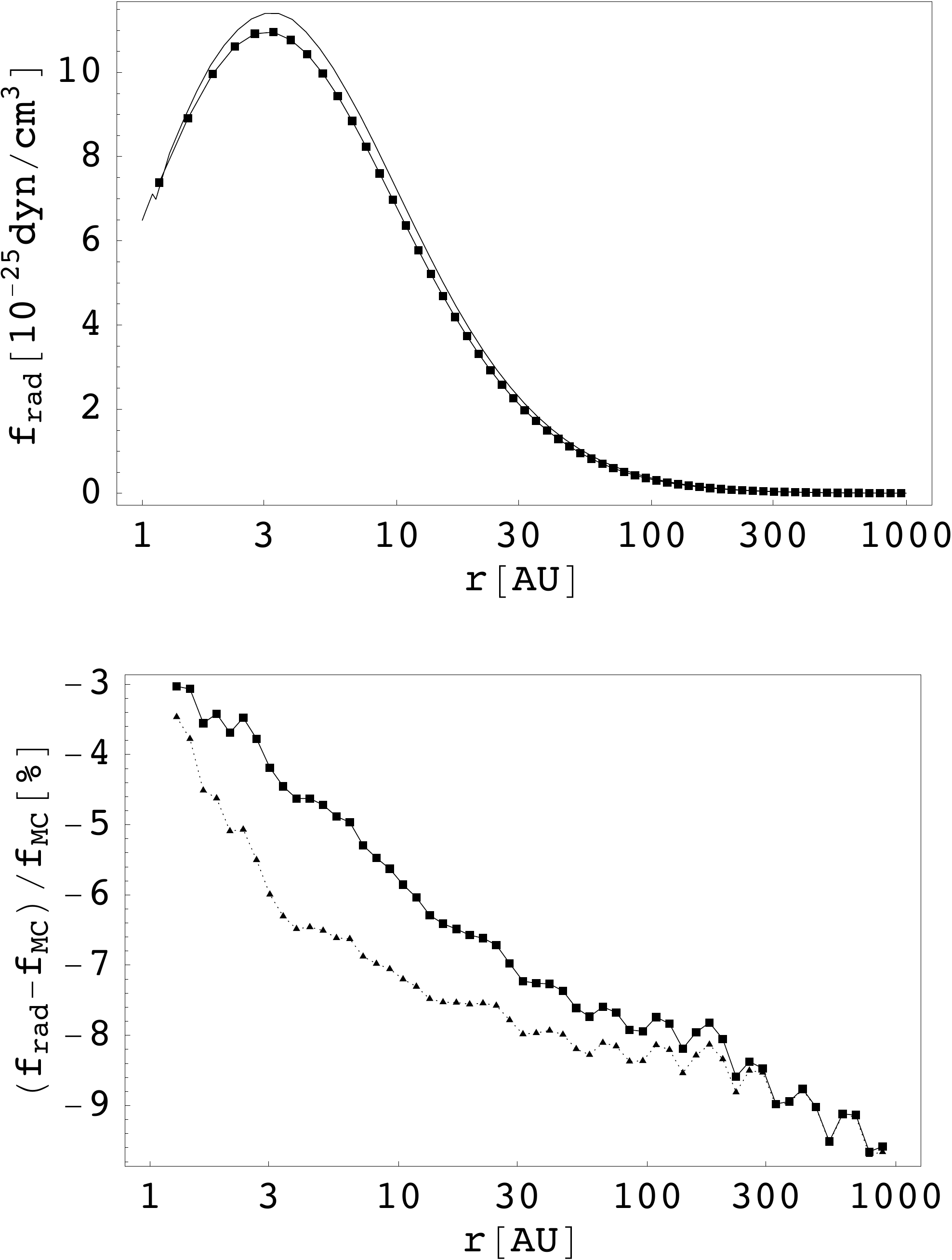}
\caption{
Radial cut through the radiative force profile at a polar angle of $\theta \approx 27\degr$ from the midplane in the
most optically thick case $\tau_{550\mathrm{nm}} = 100$. 
The profile visualizes the onset of the radiative force at the transition from the optically thin envelope to the
optically thick disk region.
\newline
Upper panel: 
Radial radiative force profile of frequency dependent irradiation plus Flux Limited Diffusion (solid line with squares) 
as well as the result from the Monte-Carlo routine (solid line).
\newline
Lower panel: 
Deviations of purely stellar (dotted line with triangles) 
and stellar plus thermal radiative force (solid line with squares) 
from the Monte-Carlo run in percent.
}
\label{fig:RadiativeForce}
\end{center}
\end{figure}

The radiative force density is, 
accordingly to
\citet{Mihalas:1984p9889},
given by
\begin{equation}
\label{eq:27}
\vec{f}_\mathrm{rad} = \rho ~ \kappa ~ \frac{\vec{F}_\mathrm{tot}}{c}.
\end{equation}
In our split radiation transport the total radiative flux $\vec{F}_\mathrm{tot}$ is given by the sum of the irradiated
stellar flux $\vec{F}_*$ from the ray-tracing routine and the diffuse component $\vec{F}$ from the FLD solver. 
The flux calculated in the FLD approximation is $\vec{F} = -D ~ \vec{\nabla} E_\mathrm{R}$ 
(see Sect.~\ref{sect:fld}). 

In discretized space the opacity and density are constant over a single grid cell and the irradiated
stellar radiative flux $\vec{F}_*$ is calculated at the cell interfaces. 
To calculate the mean radiative force inside this cell 
it is necessary to integrate the radiative force over the cell volume 
(e.g.~a simple ansatz of averaging only the stored fluxes at the interfaces towards the cell center would lead to
unphysically high radiative forces for $\tau >> 1$, 
i.e.~for the case that most of the flux is absorbed on a length scale much smaller than the grid size).
Integrating the above formula Eq.~\eqref{eq:27} over space 
(for simplicity here: an one-dimensional cartesian grid
with an uniform grid spacing of $\Delta x$) 
leads to:
\begin{eqnarray}
\label{eq:28}
\vec{f}_* &=& \frac{1}{\Delta x} \int_0^{\Delta x}\rho ~ \kappa ~ \frac{\vec{F}_*}{c} ~ dx\\
\label{eq:29}
&=& \frac{1}{c ~ \Delta x} ~ \rho ~ \kappa \int_0^{\Delta x} \vec{F}_* ~ dx
\end{eqnarray}
The flux at position $x$ inside the grid cell is given by the flux $\vec{F}_*^i$ entering the cell at the interface $i$
and the absorption of this flux along the length $x$: 
\begin{equation}
\label{eq:30}
\vec{F}_*\left(x\right) = \vec{F}_*^i e^{-\tau} \nonumber
\end{equation}
with the optical depth $\tau = \kappa ~ \rho ~ x$. 
The remaining integral yields:
\begin{eqnarray}
\label{eq:31}
\vec{f}_* &=& \frac{1}{c ~ \Delta x} ~ \rho ~ \kappa ~ \vec{F}_*^i \int_0^{\Delta x} \exp\left(-\kappa ~ \rho ~ x
\right) ~ dx\\
\label{eq:32} 
&=& - \frac{1}{c ~ \Delta x} ~ \vec{F}_*^i ~ \left(\exp\left(-\kappa ~ \rho ~ \Delta x \right) - 1 \right).
\end{eqnarray}
Finally, the mean radiative force is therefore given by 
the difference of the left and the right flux into and/or out of the cell respectively
\begin{equation}
\label{eq:33}
\vec{f}_* = - \frac{1}{c} ~ \frac{\vec{F}_*^{i+1} - \vec{F}_*^i}{\Delta x},
\end{equation}
which equals in continuous space the derivative of the radiative flux
\begin{equation}
\label{eq:34}
\vec{f}_* = - \frac{1}{c} ~ \partial_x \vec{F}_*.
\end{equation}
Combining Eqs.~\eqref{eq:27} and \eqref{eq:34} reflects the fact that without emission the radiative flux is
given by the differential equation
\begin{equation}
\label{eq:35}
\partial_x \vec{F}_* = - \rho ~ \kappa ~ \vec{F}_*.
\end{equation}
In other words,
the relationship in Eq.~\eqref{eq:35} shows that the original 
Eq.~\eqref{eq:27} and the derived 
Eq.~\eqref{eq:34} are indeed identical 
(in continuous physical space), 
whereas the left hand side of 
Eq.~\eqref{eq:35} gives the expression easily available in discretized space.
In our code we use a grid in spherical coordinates, 
thus the derivative of the stellar irradiation is given by
\begin{equation}
\label{eq:36}
\vec{f}_* = - \frac{1}{c ~ r^2} ~ \partial_r \left(r^2 \vec{F}_*\right).
\end{equation}
So the purely geometrical dilution of the flux in the radially 
outward direction
$\left(\vec{F}_*(r) \propto r^{-2}\right)$ 
does of course not contribute to the radiative force.

We compute the radiative force in the setup of the radiation benchmark test with gray and frequency dependent
irradiation as well as the thermal radiative force for the most optically thick case
and compare our results with the corresponding Monte-Carlo based run.
The result is visualized in Fig.~\ref{fig:RadiativeForce}.
The peak position is reproduced very well.
The absolute value of the peak is underestimated by 3-4\% only.
Behind the absorption peak the radiative force smoothly drops down and the relative deviations grow radially outwards,
but stay below 10\%.
The fraction of the radiative force, 
resulting from the thermal flux, 
is relatively small and is most important at and directly after the absorption peak 
(where most of the thermal radiation is emitted). 
This fraction will presumably be higher in denser environments.
At larger radii 
($r > 300$ AU) 
where the disk becomes highly optically thin for its own thermal radiation, 
the radiative force resulting from the thermal flux is negligible 
(cp.~Figs.~\ref{OpticalDepth} and \ref{fig:RadiativeForce}).

The gray approximation and the corresponding frequency dependent run show only small deviations 
$\left(< 5\%\right)$
at the outer part of the disk for radii roughly larger than 200~AU 
(after the absorption of the stellar irradiation). 
With the setup of 
\cite{Pascucci:2004p39} 
the gray approximation leads to a higher radiative force than the frequency dependent ones, 
but in general the difference of both methods depends strongly on the stellar luminosity.
The black body spectrum of 
more luminous stars will shift to higher frequencies (cp.~Fig.~\ref{Opacities}).

Further remarks, 
explanations, 
and detailed discussion of the resulting radiative force and temperature profiles are given in the following 
Subsect.~\ref{sect:remarks}.

\subsection{Remarks and Analysis}
\label{sect:remarks}
\subsubsection{The Monte-Carlo comparison code}
For comparison and interpretation of the results, 
we should mention that in the original radiation benchmark test
\citep{Pascucci:2004p39}
the different Monte-Carlo codes
themselves differ in the radial temperature profile through the midplane of the circumstellar disk in 
the most optically thick case 
$\tau_{550\mathrm{nm}} = 100$ 
by 5\% in most of the region between 1.2 to 200 AU 
and up to 15\% towards the outer border of the computational domain in the radial direction.
This means that
the deviations in the optically thin case (Fig.~\ref{Tau0001}) as well as the frequency dependent ray-tracing
part of the optically thick case (Fig.~\ref{Tau1000ab}) stay beneath the discrepancy of the different Monte-Carlo
solutions.
As expected, 
the direct stellar irradiation is determined highly accurately, 
when considering the frequency dependence. 
Therefore, 
the errors introduced by using the FLD approximation can be limited in the test throughout the irradiated regions.

The influence of the so-called photon noise in the Monte-Carlo method is illustrated in the highly optically thin regions
($|\theta| > 30\degr$ from the midplane) 
in Fig.~\ref{Tau1e+2_polar}, 
where the temperature should actually be independent for large polar angles 
(as displayed by the solid line).

\subsubsection{The Flux Limited Diffusion approximation}
\begin{figure}[ht]
\begin{center}
\includegraphics[width=0.48\textwidth]{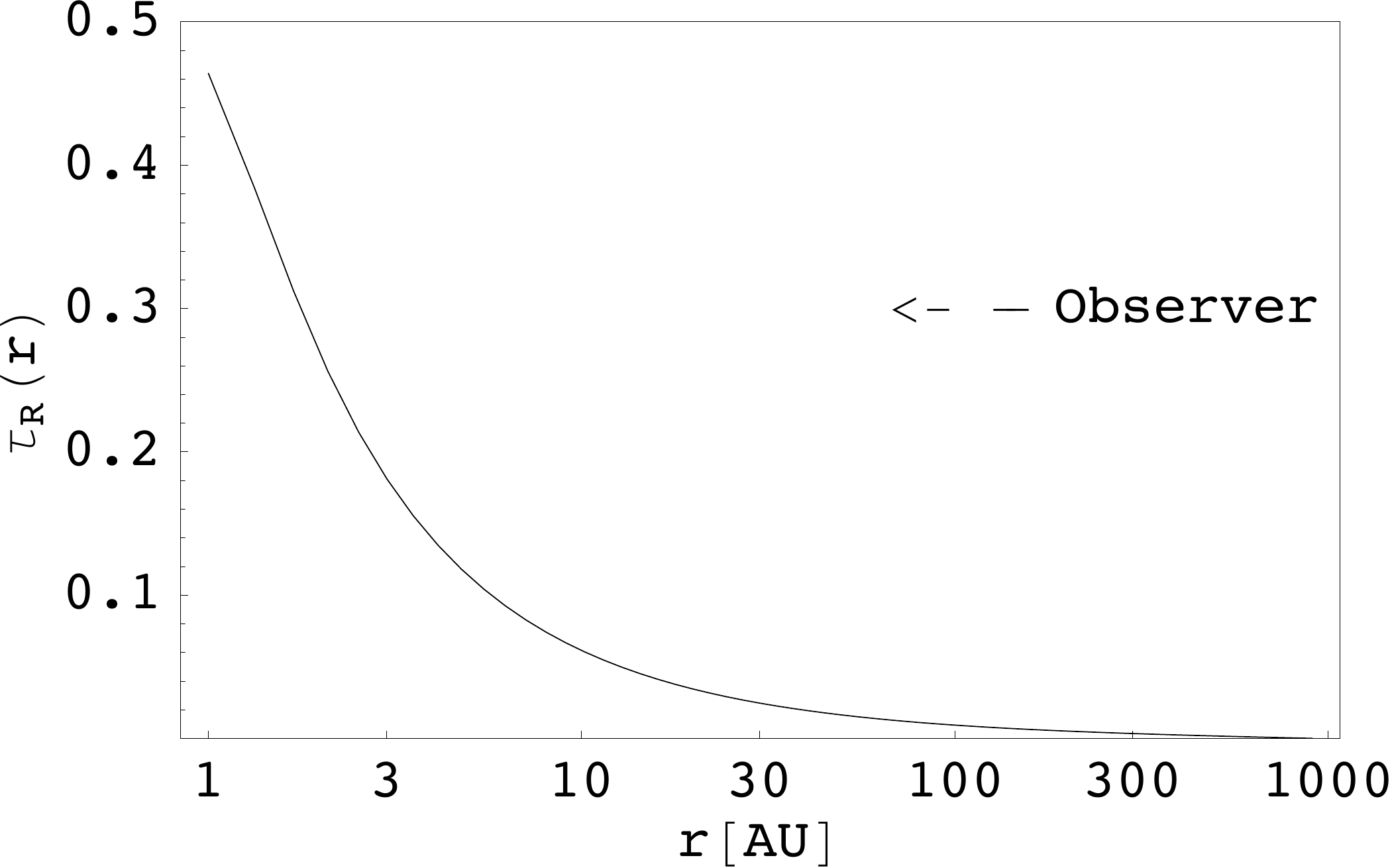}
\caption{
Radial profile of the optical depth $\tau_\mathrm{R}$ through the midplane in the most optically thick case
$\tau_{550\mathrm{nm}} = 100$ integrated from the outer edge of the disk inwards. 
}
\label{OpticalDepth}
\end{center}
\end{figure}

A special feature of the setup of the original benchmark test 
\citep{Pascucci:2004p39} 
is the fact that even in the most optically thick case $\tau_{550\mathrm{nm}} = 100$, 
which is defined for a wavelength of $550$ nm, 
the pre-described disk is locally optically thin for the radiation from thermal dust emission. 
Integrating the corresponding local optical depth 
$\tau_\mathrm{R}\left(r\right) = \kappa_\mathrm{R}\left(T\right) ~ \rho\left(r\right) ~ \Delta r$ 
from the outer edge of the disk through the midplane towards the center yields a final
optical depth of $\tau_\mathrm{R} \approx 0.5$, 
see Fig.~\ref{OpticalDepth}.

The plot shows clearly the low optical depth for the thermal component of the radiation field 
especially in the outer part of the disk, 
which results in an underestimation of the temperature in the transition region at roughly 
$r \approx$~200~AU due to the overestimation of the radiative flux in the outward direction by the FLD approximation. 

The FLD approximation is known to be valid in the most optically thin (free-streaming limit) 
as well as in the most optically thick (diffusion limit) regions only. 
The apparent yet surprisingly good agreement between the Monte-Carlo based runs 
and the herein proposed radiation transport module also in the intermediate region of the flared disk atmosphere 
(see Fig.~\ref{Tau1e+2_polar}) is due to the newly implemented direct irradiation routine
which yields the correct flux and depth of penetration for the different frequency bins of the stellar irradiation
spectrum (see Sect.~\ref{sect:irradiation_thick}). 
The slight underestimation of the temperature at $r \approx 200 \mbox{ AU}$ 
(see Fig.~\ref{Tau1e+2_Irradiation+FLD_radial}) 
in the disk midplane in the most optically thick case is most likely a
result of an intermediate region 
(transition from optically thick to optically thin) in the outward radial direction, 
which is shielded from the direct irradiation and is not in good agreement with the FLD approximation.

\subsubsection{The frequency dependence}
Approximating the frequency dependence of the stellar spectrum by gray 
(frequency averaged)
Planck mean opacities results in an
overestimation of the optical depth in the infrared part
and an underestimation of the absorption in the ultraviolet part of the stellar spectrum (see Fig.~\ref{Opacities}). 
This leads to an under- or overestimation of the radiative force onto dust grains at the first absorption peak
depending on the stellar luminosity 
due to less absorption of the most energetic UV photons 
or stronger absorption of the photons at the peak of the stellar black body spectrum.
Secondly, 
the gray approximation results in an underestimation of the resulting temperature deeply inside the disk 
due to absorption of the infrared photons already at inner disk radii.
The quantitative amount of the resulting deviations depends in general on the specific stellar and dust properties. 
Due to the steep decay of the stellar black body spectrum at high frequencies, 
the difference of the gray and frequency dependent radiative force turns out to be very small in this specific setup.
On the other hand,
consideration of the frequency dependence is essential to limit the deviations in the resulting temperature profile to
less than 11.1\% 
(compared to 38.4\% for gray irradiation plus FLD, see Fig.~\ref{Tau1e+2_Irradiation+FLD_radial}). 
These issues are well illustrated in Fig.~\ref{Opacities}.
The figure shows the frequency dependent opacities of
\citet{Draine:1984p594}, 
the approximated frequency averaged value of the Planck mean opacity regarding the stellar effective surface temperature
as well as the black body spectrum of the central star 
(to visualize the amount of radiative flux which is emitted per frequency bin). 
Each dot in the figure marks the mid-frequency of the correspondingly chosen frequency bin. 
The effects on the resulting temperature 
profile and radiative force cannot be generalized easily.
They depend on the underlying dust model and strongly on the properties of the central star, 
which yield a shift of the peak position of the black body spectrum in Fig.~\ref{Opacities} according to Wien's
displacement law $\nu_\mathrm{peak} \propto T$. 
In the specific setup of 
\citet{Pascucci:2004p39} 
the Planck mean opacity at the black body peak position is higher than the frequency dependent ones, 
leading to a slightly higher radiative force. 
The strong overestimation of the opacity in the infrared regime leads to the huge discrepancy of the gray approximation
in the radial temperature profile through the disk.

\section{Parallel performance of the approximate radiation transport module}
\label{sect:performance}
The parallelization of the radiation transport scheme and the GMRES solver 
(see Sect.~\ref{sect:gmres} for details of how the solver works) 
are taken care of by the PETSc library 
(Portable, Extensible Toolkit for Scientific computation), 
see also \cite{petsc-efficient}. 
To test the parallel speedup of the implemented radiation transport module we
perform two tests with an extended version of the circumstellar disk setup introduced in Sect.~\ref{sect:setup}.

We adopt the most optically thick setup for $\tau_{550\mbox{nm}} = 100$ and expand it to three dimensions assuming
axial symmetry. 
All runs include frequency dependent irradiation and Flux Limited Diffusion.
The tests run for 10 main iterations, 
which consume the main computational effort for the approximate solver 
(later on, near equilibrium, the internal iterations needed decrease strongly). 
The number of internal iterations of the approximate
implicit solver is fixed to 100 to guarantee the same amount of
computation in all runs during this benchmark test.
Due to the parallel Block-Jacobian pre-conditioner, 
the number of needed internal iterations 
(for a specified accuracy) 
normally increases with increasing number of processors. 
The precise value for the increase is strongly problem dependent 
(B.~F.~Smith, member of the development team of the PETSc library, private communication).

The parallel domain 
(the linear system of equations) 
is split in the azimuthal and polar direction only, which insures good speedup and efficiency.
Decomposing the domain in the radial direction would decrease the parallel
performance due to the fact that in the ray-tracing routine it is necessary to
compute and therefore communicate the flux from the central sink cell to the outer boundary 
from the inside outward.
Since the knowledge of the flux at the inner cell interface is needed to
compute the flux at the outer interface, 
this method is hardly parallelizable as a domain decomposition.

The measured times $t_2$ to $t_n$ ($n$ is the number of processors) represent the wall clock time per main iteration per
grid cell without the non-recurring initialization and finalization of the code. 
We perform runs with 2 up to 64
processors due to the fact that single job submission is not available on the cluster we use. 
Cases, 
in which the local cache size would exceed the parallel decomposed problem size, 
are not taken into account,
i.e.~no misleading super linear speedup for high number of used processors is shown here. 
The speedup $S$ is
calculated as the ratio of the `serial' run time compared to the wall clock time used by the parallel run: 
$S=t_2 / t_n$. 
The efficiency $E$ is determined via 
$E = t_2 / (t_n ~ n) = S/n$. 
Each run for a specific grid and a specific number of processors is
performed three times and averaged afterwards, 
but the differences of the resulting run times are negligible.

All tests are performed on a 64-Bit Opteron cluster consisting of 80 nodes with two CPUs each.

\subsection{The constant grid test}
\begin{figure}[ht]
\begin{center}
\includegraphics[width=0.48\textwidth]{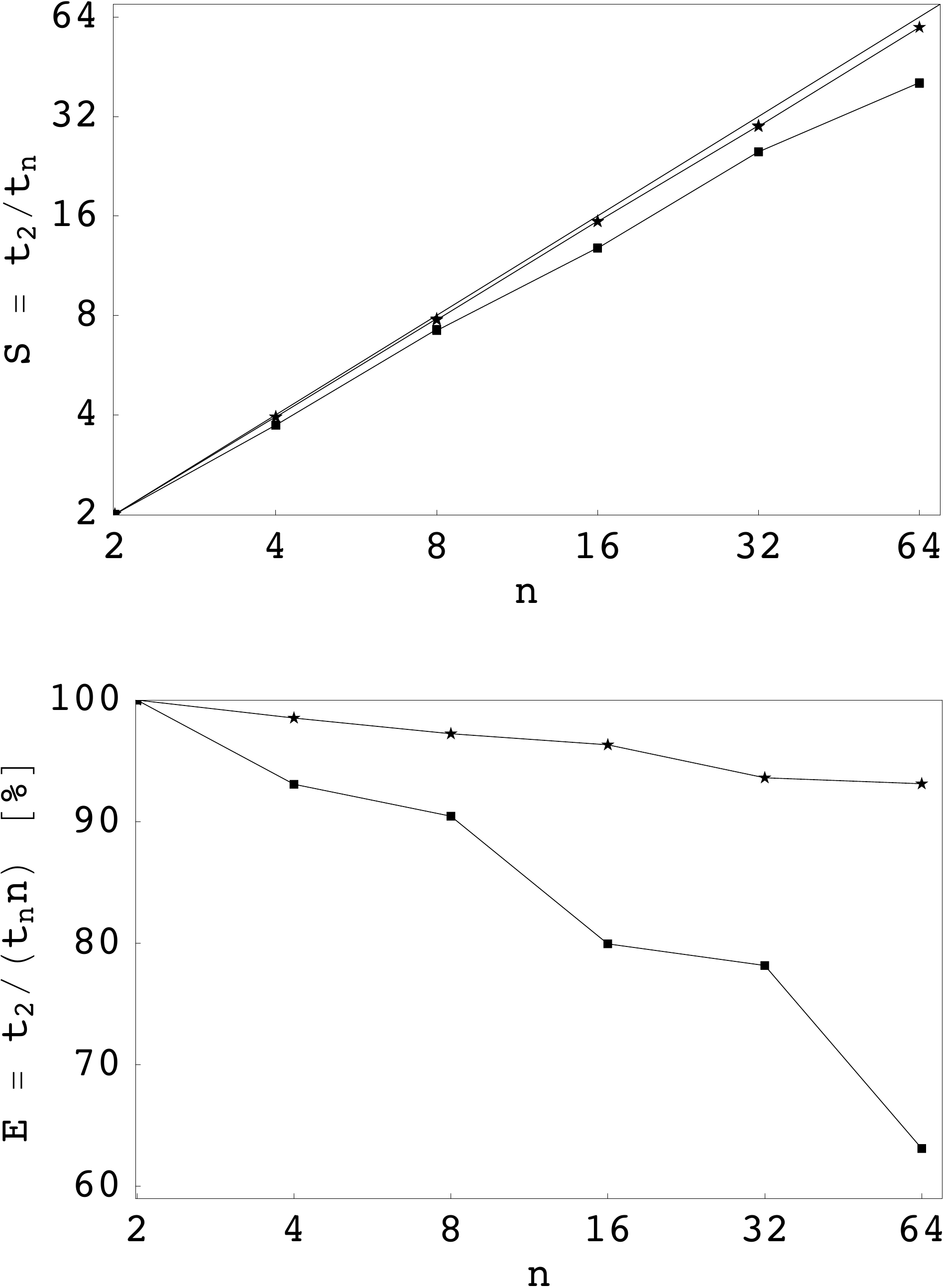}
\caption{
Speedup factors $S=t_2 / t_n$ (upper panel)
and corresponding efficiencies $E=t_2 / \left(t_n ~ n\right) = S/n$ (lower panel)
for a fixed (squares) 
and a growing problem size (stars). 
}
\label{Performance}
\end{center}
\end{figure}

The runs during this test are performed on a grid consisting of $64 \mbox{ x } 64 \mbox{ x } 256$ grid
cells. 
Each processor covers therefore a $\left(64 \mbox{ x } 64 \mbox{ x } 256\right) / n$ subdomain, 
depending on the number of processors used. 
This test shows the speedup one can gain when running a fixed problem on more and more processors. 
Therefore, 
the parallel efficiency declines stronger than in the following growing grid test 
due to the fact that with the usage of an increasing number of processors one lower the amount of computation and
increase the amount of communication per single CPU. 
This means that the granularity (ratio of computation to communication) of the parallel problem drops
strongly with an increasing number of processors.
The resulting speedup factors and efficiencies of this test are shown in Fig.~\ref{Performance}.

\subsection{The growing grid test}
The runs during this test are performed on a grid consisting of 32,768$*n$ grid cells 
(e.g.~a $64 \mbox{ x } 64 \mbox{ x } 8n$ grid), 
where $n$ is again the number of processors used. 
Each processor covers therefore a  subdomain containing 32,768 grid cells respectively during all runs, 
independent of the number of processors used.
This test shows the efficiency achievable when using more processors to run a bigger problem.
This is a realistic setup for our three-dimensional radiative hydrodynamics studies of collapsing massive
pre-stellar cores, 
which are at the current limit of our clusters. 
The resulting speedup factors and efficiencies are shown in Fig.~\ref{Performance}.

\subsection{Parallel performance results}
The constant grid test shows a clear speedup for the fixed problem, 
so it is simply possible to compute a fixed problem faster by using more processors. 
The growing grid case shows a high efficiency of more than 95\% during all runs. 
This speedup seems to be higher than the speedup of any actual freely available three-dimensional hydrodynamics code for
spherical grids 
(well known in the astrophysical community, e.g.~Pluto2, Flash2.5, and Zeus-MP1.5), 
which we tested during the summer 2007 in an accretion disk setup in hydrostatic equilibrium. 
In this sense, 
the resulting parallel speedup is high enough for the planned integration of this radiation transfer module into a
(magneto-) hydrodynamics framework.

Finally, 
we can recommend the usage of such an implicit modern KSP solver method leading to fast convergence of the radiative
diffusion problem 
(at least for the setup discussed)
while simultaneously offering high parallel efficiency. 
Admittedly,
attention should be paid to the general fact that such an approximate solver method strongly depends on the
physical problem at hand as well as on the specified accuracy or abort criterion.

\section{Radiative hydrodynamics shock tests}
\label{sect:radiativeshock}
To test the approximate radiation transport scheme also in dynamical interaction
with a streaming fluid, 
we perform two standard radiative shock tests.
We adopt the setup of the supercritical and subcritical radiative shock tests for the VISPHOT
code in \citet{Ensman:1994p10128}.
These radiative shock tests were already repeated in tests of the TITAN code 
\citep{Sincell:1999p10169, Sincell:1999p10170}, 
ZEUS-2D
\citep{Turner:2001p1152},
as well as ZEUS-MP 
\citep{Hayes:2006p1133}.
Analytic approximations for this kind of problem are given by
\citet{ZelDovich:1967p10174}.

\begin{figure*}[t]
\begin{center}
\includegraphics[width=0.95\textwidth]{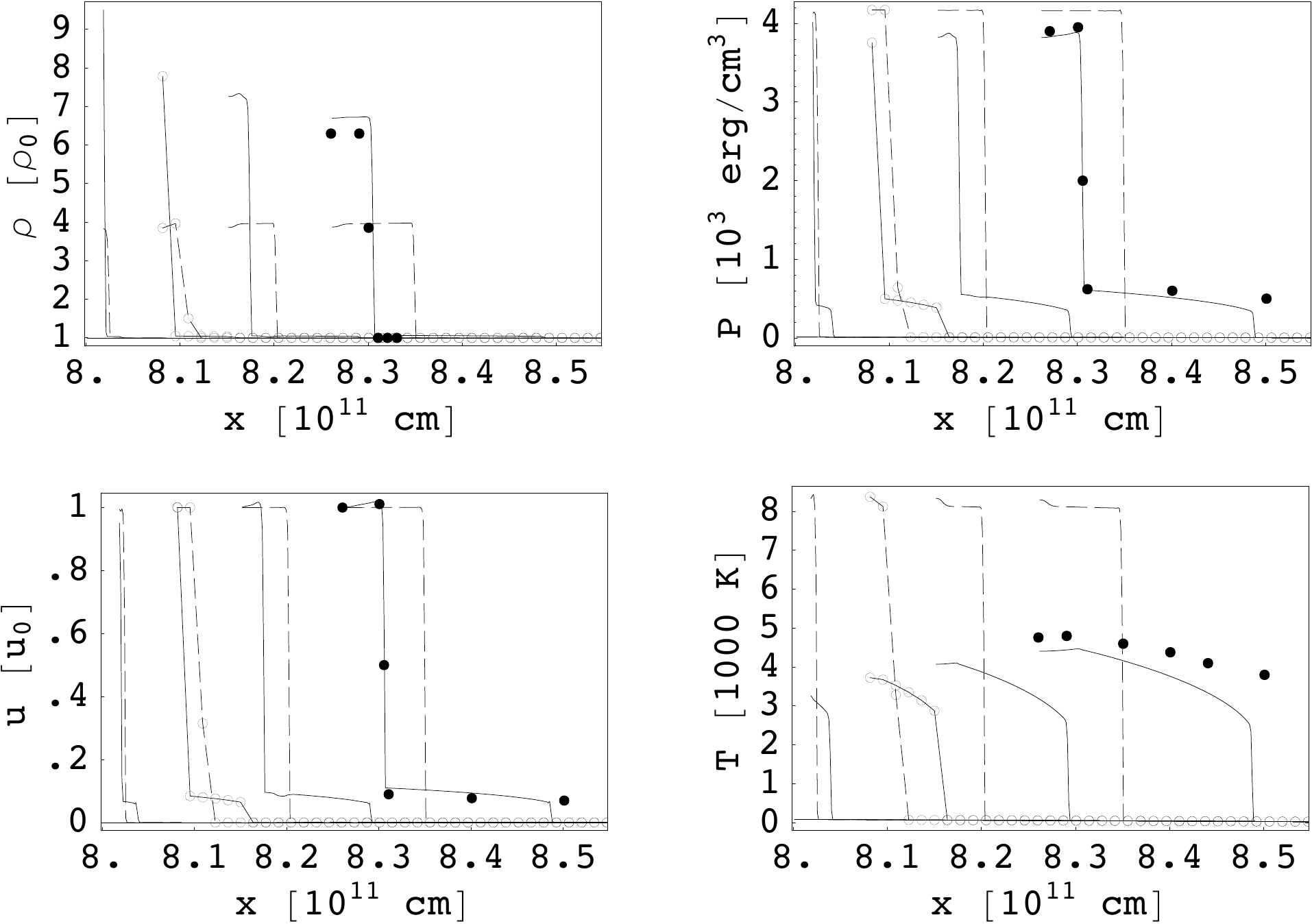}
\caption{
Radiative supercritical shock.
Resulting density, 
pressure, 
velocity,
and temperature distributions at four different snapshots in time.
Dashed lines represent the adiabatic runs, 
solid lines the radiative ones.
The time snapshots are taken 
(from left to right) 
at 860 s, 4,000 s, 7,500 s, and 13,000 s after launching. 
Mostly horizontal lines at the lower border of the graphics refer to the initial setup. 
The snapshots at 4000 s are additionally marked by open circles for every 
$10^\mathrm{th}$ grid cell to illustrate the resolution used. 
The spatial axes are retranslated into the non-moving frame used in the visualization by 
\protect\citet{Ensman:1994p10128}  
for the sake of
comparison.
The filled dots in the four profiles at 13,000 s represent characteristic sample points in the two-temperature
approach read off Figs.~10~-~12 in the original study by
\protect\citet{Ensman:1994p10128}.
}
\label{RadiativeSupercriticalShock}
\end{center}
\end{figure*}

To follow the motion of the gas, 
we solve
(additionally to the radiation transport equations of Sect.~\ref{sect:theory}) 
the equations of compressible hydrodynamics in conservative form
\begin{eqnarray}
\label{eq:37}
\partial_t \rho + \vec{\nabla} \cdot \left(\rho ~ \vec{u}\right)                 &=& 0 \\
\partial_t \left(\rho ~ \vec{u}\right) + \vec{\nabla} \left(\rho ~ \vec{u} ~ \vec{u}\right) &=& - \vec{\nabla} ~ P +
\rho ~
\vec{a}\\ 
\partial_t E + \vec{\nabla} \cdot \left(\left(E + P\right) ~ \vec{u}\right)      &=& \rho ~ \vec{u} \cdot \vec{a}
\end{eqnarray}
with the acceleration source term from the radiation transport
\begin{equation}
\vec{a} 
= - \frac{\vec{\nabla} \cdot \vec{F_*}}{\rho ~ c} \vec{e}_r - \kappa_\mathrm{R} \frac{D ~ \vec{\nabla} E_\mathrm{R}}{c},
\end{equation}
derived in Sect.~\ref{sect:radiativeforce}.
The evolution of 
the gas density $\rho$, 
the gas velocity $\vec{u}$, 
the thermal pressure $P$, 
and the total energy $E$ 
is computed using the open source magneto-hydrodynamics code Pluto3 
\citep{Mignone:2007p3421}. 
Pluto is a high-order Godunov solver code, 
i.e.~it uses a shock capturing Riemann solver within a conservative finite volume scheme. 
Our default configuration consists further of 
a Harten-Lax-VanLeer (hll) Riemann solver 
and a so-called ``minmod'' flux limiter 
using piecewise linear interpolation (plm) 
and a Runge-Kutta 2 (RK2) time integration 
\citep[predictor-corrector-method, compare e.g.][]{vanLeer:1979p5193}. 
Therefore the total difference scheme is $2^\mathrm{nd}$ order accurate in time and space.

The test setup describes a piston moving with supersonic velocity through an initially uniform, cold gas. 
The one-dimensional domain of the grid associated covers a distance of length 
$l_0=7*10^{10}\mbox{ cm}$.
The iso-density in the domain is fixed to $\rho_0 = 7.78*10^{-10} \mbox{ g cm}^{-3}$.
For testing purposes, 
the gas is set to be completely ionized,
thus the mean molecular weight is 
$\mu = 0.5$.
An ideal equation of state is used with $\gamma = 5/3$. 
The opacity is fixed to a constant value of 
$\kappa = 0.4
\mbox{ cm}^2 \mbox{ g}^{-1}$. 
The initial temperature drops linearly from 85~K at the starting position
of the piston to 10~K at the outer boundary.
The velocity $u_0 > c_\mathrm{s}$ of the piston is used to determine the strength of the shock. 
While the piston moves through the domain the radiative energy from the
shocked gas will stream upwards leading to 
a pre-heating, 
a pre-acceleration, 
as well as a pre-compression of the gas directly in front of the shock.
If the temperature in this pre-heated region stays below the temperature of the shocked gas, 
it is called a subcritical radiative shock.
If the temperature in the pre-heated region equals the temperature of the shocked gas, 
it is called a supercritical radiative shock. 
The smallest piston velocity leading to a supercritical shock defines the critical velocity $u_\mathrm{c}$.

\begin{figure*}[t]
\begin{center}
\includegraphics[width=0.95\textwidth]{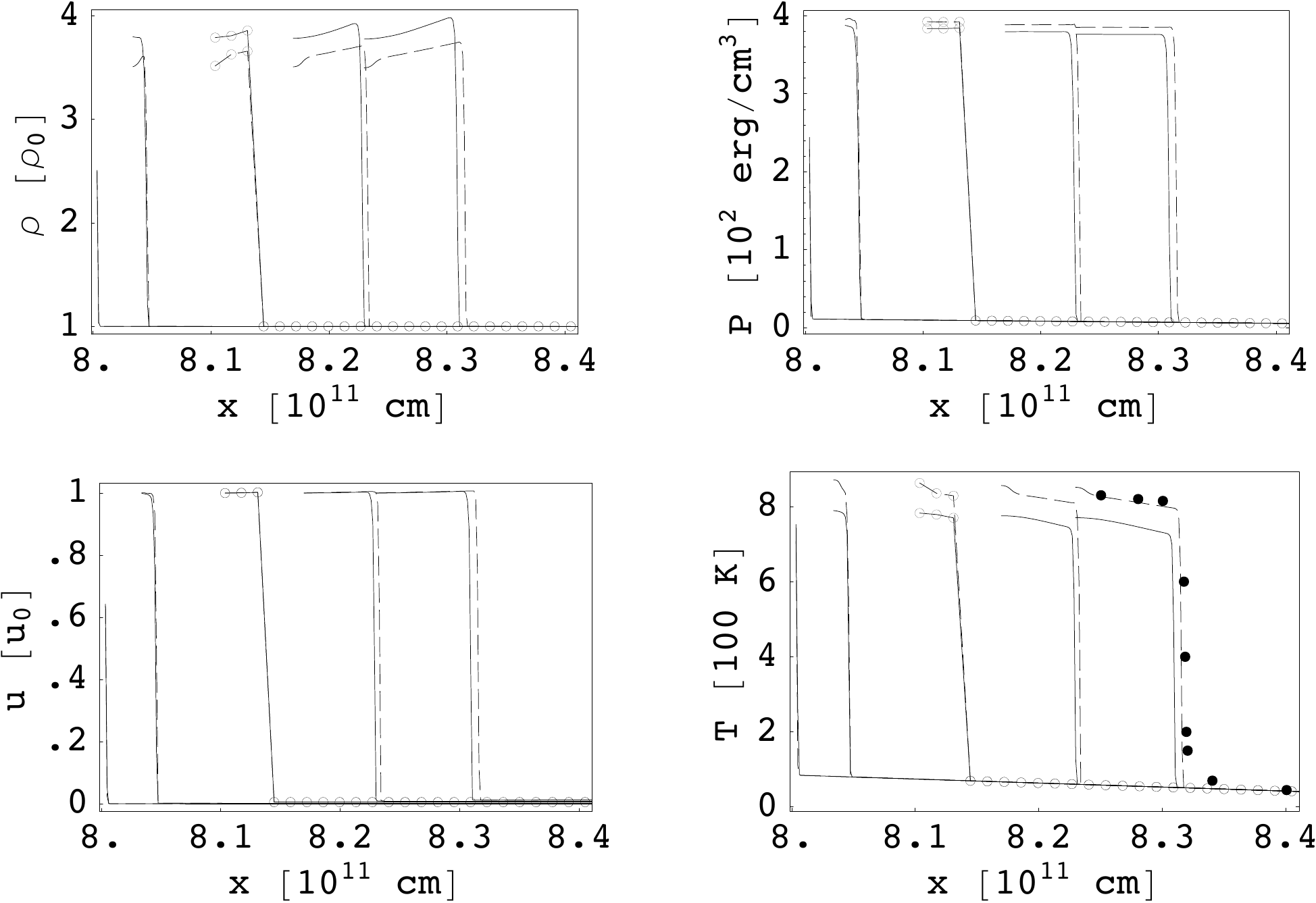}
\caption{
Radiative subcritical shock.
Resulting density, 
pressure, 
velocity, 
and temperature distributions at five different snapshots in time.
Dashed lines represent the adiabatic runs, 
solid lines the radiative ones.
The time snapshots are taken (from left to right) 
at 350 s, 5,400 s, 17,000 s, 28,000 s, and 38,000 s after launching. 
The snapshot at 350 s mostly refers to the initial setup. 
The snapshots at 5,400 s are additionally marked by open circles for every $10^\mathrm{th}$ grid cell to illustrate the
resolution used. 
The spatial axes are retranslated into the non-moving frame used in the visualization by 
\protect\citet{Ensman:1994p10128} 
for the sake of comparison.
The filled dots in the temperature profile at 38,000 s represent characteristic sample points in the
two-temperature approach read off Fig.~8 in the original study by 
\protect\citet{Ensman:1994p10128}.
}
\label{RadiativeSubcriticalShock}
\end{center}
\end{figure*}

\citet{Ensman:1994p10128} used a Lagrangian grid moving with the piston velocity.
This setup is translated into an Eulerian grid by setting the initial velocity in
the whole domain as well as the permanent velocity at the outer boundary to the
negative of the piston velocity $u_0$, 
compressing the gas at the inner reflective boundary, 
which represents the moving piston. 
In the visualization of our results 
(see Figs.~\ref{RadiativeSupercriticalShock} and \ref{RadiativeSubcriticalShock})
the spatial axes are retranslated into the non-moving frame used in the visualization by 
\citet{Ensman:1994p10128} 
for the sake of comparison. 
The spherical coordinate system used at large radii (to achieve a planar geometry) by
\citet{Ensman:1994p10128}
is translated into cartesian coordinates.
We use 512 uniform grid cells to cover the spatial extent of the grid.
These grid adjustments were also used in the test of the ZEUS-MP code 
\citep{Hayes:2006p1133}.

In this radiative shock test setup we are able to check the dynamical behaviour of the radiation module and the
hydrodynamics. 
On the other hand, 
it implies a much easier treatment of radiation transport 
(due to the fact that the optical depth in front of the shock is practically constant) 
than the prior static but frequency dependent benchmark test of 
\citet{Pascucci:2004p39}.
Therefore, 
only the FLD routine 
(as described in Sect.~\ref{sect:fld}) 
is needed to perform the problem runs.

We study this shock scenario in purely adiabatic as well as radiative hydrodynamics simulations.

\subsection{Radiative supercritical shock}
Higher velocity of the piston results in a stronger pre-heating of the gas directly in front of the shocked gas.
When a piston velocity higher than the critical velocity $u_\mathrm{c}$ is used, 
the peak temperature of the pre-heated region 
(which is equal to the temperature of the shocked gas) 
will not increase any more, 
but the enlargement of the pre-heated region will extend further out.
The gas temperature of the pre-heated region will presumably be in equilibrium with the radiation temperature for
the most part of the domain. 
At one point of the sloping tail of the temperature distribution in the upstream direction both
temperatures will differ. 
In an equilibrium diffusion (also called one-temperature diffusion) code like ours the temperature will decline at this
point sharply 
(see also Fig.~15 in \citet{Ensman:1994p10128}, which compares an adiabatic, an equilibrium,
a non-equilibrium and a full radiation transport method for this setup).
The reason for this sharp decline is 
(according to \citet{Ensman:1994p10128})
that the radiation can only penetrate through the cold environment, 
if it heats this environment up to the equilibrium temperature.
Once the radiation energy is absorbed the radiative flux is zero.
In a non-equilibrium radiation transport method the radiation energy can penetrate further into the environment simply
by heating it to less than the equilibrium temperature.

We choose the piston velocity to be the one used originally by
\citet{Ensman:1994p10128},
$u_0 = 20 \mbox{ km} \mbox{ s}^{-1}$.
In some of the subsequent tests mentioned above a slightly lower piston velocity 
$\left(u_0 = 16 \mbox{ km} \mbox{ s}^{-1}\right)$
for the supercritical radiative shock was used by the authors.

Fig.~\ref{RadiativeSupercriticalShock} displays the resulting 
density, 
pressure, 
velocity,
and temperature distributions
for four different times 
(same as Figs.~10~-~12 in \citet{Ensman:1994p10128}) 
as well as the initial setup. 
The analytic limit of a maximum jump in density by four in adiabatic shocks is reproduced. 
The effects of 
pre-heating, 
pre-acceleration, 
and pre-compression are clearly visible. 
The position of the peak temperature and the sharp declines in the density, 
the pressure, 
and the velocity 
at the location of the shock front
fit the one by \citet{Ensman:1994p10128}. 
That means, 
the propagation speed of the shock is reproduced correctly.
Although the non-equilibrium gas temperature spike, 
the maximum temperature, 
cannot be reproduced with an equilibrium radiation transport method 
(the width of this gas temperature spike would always be less than a mean free path across), 
the value of the peak temperature is in good agreement with the radiation temperature of the corresponding
non-equilibrium run by 
\citet{Ensman:1994p10128}
aside from minor geometrical effects of the modified coordinate system as already discussed in \citet{Hayes:2006p1133}. 
Furthermore, 
the equilibrium temperature distribution resembles the temperature distribution of the non-equilibrium run by
\citet{Ensman:1994p10128} 
in regions, 
where the gas temperature equals the radiation temperature 
(which is the most part of the domain in such a supercritical shock). 
As expected,
the temperature 
profile calculated by our equilibrium diffusion method
sharply declines at the boundary to the non-equilibrium region predicted by
\citet{Ensman:1994p10128}.
The outermost data points by \citet{Ensman:1994p10128} in 
the pressure, 
the velocity, 
and the temperature
distributions in Fig.~\ref{RadiativeSupercriticalShock} display the long-range tail of the pre-heated region, 
if using a non-equilibrium radiation transport method.
Summing up,
the deviations of the equilibrium temperature method from a non-equilibrium approach stay below 10\% in the pre- and
post-shock regions.
In this dynamic diffusion shock test,
the outermost tail of the pre-heated region, 
where the gas and the radiation temperature is not in equilibrium, 
can naturally not be observed in an equilibrium approach.
In the context of massive star formation, 
\citet{Krumholz:2007p855} already show that massive star formation is generally in the static diffusion limit. 

\subsection{Radiative subcritical shock}
In a radiative subcritical shock ($c_\mathrm{s} < u_0 < u_\mathrm{c}$) the gas in front of the shock
region is heated to a temperature lower than the temperature of the
shocked gas region only.
In a non-equilibrium radiation transport method the gas temperature sharply
drops down 
(after the narrow temperature spike similar to the supercritical shock), 
whereas the radiation energy penetrates the cold unshocked gas region and declines more smoothly 
(see also Fig.~8 in \citet{Ensman:1994p10128}).

An equilibrium radiation transport is expected to first ignore the narrow gas temperature spike.
Secondly, it drops down sharply when arriving the region, 
where the gas and radiation temperature are out of equilibrium.
In other words,
the equilibrium radiation transport resembles the decline of the gas temperature distribution.

We choose the piston velocity to be the one used originally by
\citet{Ensman:1994p10128},
$u_0 = 6 \mbox{ km} \mbox{ s}^{-1}$.
Fig.~\ref{RadiativeSubcriticalShock} displays the resulting density,
pressure, 
velocity, 
and temperature distributions for five different times 
(same as Fig.~8 in \citet{Ensman:1994p10128}).
Given the constraints of the equilibrium radiation transport discussed above
the resulting distributions fully satisfy the predictions.

\section{Conclusions \& Outlook}
\label{sect:conclusions}
We introduced a fast and accurate radiation transport module,
feasible for prospective usage in hydrodynamics and magneto-hydrodynamics simulations of circumstellar environments.
The basic physical principle of the proposed radiation transport scheme is 
the splitting of the full radiative transfer problem 
into an accurate first order ray-tracing component 
and an approximate treatment of second order effects
\citep{Wolfire:1986p562, Murray:1994p9750, Edgar:2003p6}. 
The method combines thereby the advantages of 
the Flux Limited Diffusion approximation (speed) 
and a first order frequency dependent ray-tracing routine (accuracy). 
The thermal radiation by dust grains is solved in the 
static diffusion limit via a FLD solver.
The FLD approximation yields the correct solution in the
optically thin (free-streaming) 
and optically thick (diffusion) limit. 
But in the transition regions, 
where also the most important feedback due to stellar radiative forces occurs, 
the FLD approximation is not valid anymore. 
This first impact of stellar irradiation is therefore accurately solved in the ray-tracing routine.

Our implementation of the FLD method uses a MPI parallelized implicit GMRES restarted solver 
yielding high speedup factors comparable to modern hydrodynamics solvers. 
The ray-tracing routine, 
which calculates the distribution of stellar irradiation, 
benefits from the usage of spherical coordinates, 
in which the outgoing stellar rays are aligned with the radial coordinate axis. 
Expanding this first order ray-tracing to frequency dependent irradiation is therefore convenient 
and results in higher accuracy of the temperature slope deeply inside the disk due to lower optical depth
of the infrared part of the stellar spectrum. 
A minor flaw of the proposed radiative transfer module is the neglecting of scattering. 
But in the radiation benchmark test scattering has only a marginal influence on the
resulting temperature distribution: 
Scattering would increase the temperature in the irradiated parts 
(up to an optical depth of about unity) 
due to higher extinction 
from about 2\% in the optically thin envelope 
up to a maximum of 19\% in the optically thick inner rim of the disks midplane. 
The more effectively shielded outer regions of the disk would be about 10\% cooler. 
Secondly, 
for more massive and luminous stars the effect of scattering will decrease due to stronger forward scattering of the
most energetic UV photons, 
which is automatically included in our frequency dependent first order ray-tracing routine. 
\citet{RowanRobinson:1980p3678} stated that neglecting the effect of scattering except for the first scattering of the
starlight yield a good approximation. 
Another flaw is the frequency averaged instead of frequency dependent re-emission of once absorbed stellar photons, 
i.e.~the expansion of the ray-tracing routine to second and higher order would result in a more and more realistic 
(but also computationally more expensive) 
radiative transfer.

We tested the approximate radiation transport scheme for hydrodynamics 
within the setup of a well established standard benchmark test for 
(Monte-Carlo and finite differences based) 
radiative transfer codes \citep{Pascucci:2004p39}, 
consisting of a central star, 
a circumstellar flared disk, 
and an envelope. 
The deviations of the resulting temperature slopes stay below a maximum of 11.1\% compared to a frequency dependent
Monte-Carlo based radiation transport method, which requires much more CPU time. 
This deviation in temperature will be reflected in the associated hydrodynamics in roughly 6\% deviation in the
corresponding sound speed $c_\mathrm{s} \propto \sqrt{T}$. 
Neglecting the frequency dependence of the stellar irradiation however results in significant higher
deviations up to 38.4\% in the temperature distribution. 
Also the computed (frequency dependent) radiative force onto
dust grains is in good agreement 
(below 3\% at the inner rim 
and below 5\% throughout the first absorption peak 
up to 10\% relative deviation at the outer boundary, 
where the radiative force gets negligible) 
with the comparison radiation transport method likely due to the usage of direct stellar ray-tracing feedback. 
The overall resulting deviations potentially passed to the hydrodynamics solver are indeed smaller 
than the lack of knowledge of the appropriate opacities. 
Summing up, 
the accuracy of the proposed radiation transport method legitimates the approximations needed for adequate speed in
combination with hydrodynamics.

In shock simulations, 
we show that the one-temperature approach, which is valid in the static diffusion limit,
is sufficiently accurate for the astrophysical applications we have in mind. 
\citet{Krumholz:2007p855} already show that massive star formation is generally in the static diffusion limit.
Nevertheless an extension to a two-temperature approach is straight forward, 
if desired for other astrophysical applications in the future.

Furthermore, 
we performed parallel performance tests of the radiative transfer module with different grid and parallelization setups. 
The parallel speedup of the solver imposes no restriction on large scale radiative (magneto-)
hydrodynamics simulations.

Therefore, 
we conclude that frequency dependent radiative feedback in hydrodynamics studies of the formation of a
massive star becomes feasible with the herein proposed approximate radiation transport method. 
Hydrodynamics simulations of monolithic pre-stellar core collapse models in massive star formation, 
including direct frequency dependent stellar feedback, 
is now one of our primary goals for this radiation transport module. 
Secondly, 
three-dimensional ideal and non-ideal MHD simulations of the magneto-rotational instability (MRI) in
proto-planetary accretion disks with radiative cooling are intended
\citep{Kuiper:2009p4092}.
Thirdly, 
the formation of circumstellar disks and the evolution of their morphology in terms of potential planet
formation can be studied also around low-mass stars. 
The algorithm is fully implemented in our version of the magneto-hydrodynamics code Pluto 
(including high-order Godunov methods for hydrodynamics, 
full implementation of the viscosity stress tensor,
and additionally external forces, see \citet{Mignone:2007p3421}).

\begin{acknowledgements}
This research has been supported by the 
International Max-Planck Research School for Astronomy and Cosmic Physics at the University of Heidelberg (IMPRS-HD).
Special thanks goes to Fabian Janssen for benchmarking different hydrodynamics codes on the PIA cluster during summer 2007. 
R.~Kuiper would like to thank 
Gerhard Hoffmann for his permanent helpful suggestions and the discussions about numerical details, 
especially the parallelization, 
of the developed radiation transport module 
as well as J\"urgen Steinacker for helpful conversation about the radiation benchmark test of 
\citet{Pascucci:2004p39}. 
Author R.~Kuiper further thanks A.~Mignone 
for the main development of the open source magneto-hydrodynamics code Pluto
as well as his great support in code specific questions. 
Authors H.~Klahr, C.~P.~Dullemond and W.~Kley have been supported in part by the 
Deutsche Forschungsgemeinschaft DFG through 
grant DFG Forschergruppe 759 ``The Formation of Planets: The Critical First Growth Phase''.
Parallel performance tests have been performed on the rio cluster of the Max Planck computing center in Garching.
\end{acknowledgements}

\bibliographystyle{aa}
\bibliography{12355}

\end{document}